\documentclass[12pt]{article}

\usepackage{amsmath,amsfonts,amssymb,amsbsy}
\usepackage{graphicx}

\usepackage[colorlinks=true,linkcolor=black,citecolor=blue,urlcolor=blue,filecolor=blue]{hyperref}

\usepackage{natbib}
\usepackage{fancyvrb}
\usepackage{microtype}

\usepackage{algorithm}
\usepackage{algorithmic}
\usepackage{mathtools}

\usepackage{booktabs}
\usepackage{soul} 
\usepackage{url}
\usepackage{xcolor}
\usepackage{footnote}
\usepackage{caption}
\usepackage{multirow}
\usepackage{placeins}
\usepackage[toc,page]{appendix}

\usepackage{accents}

\usepackage{geometry} 
\usepackage{sectsty} 
\usepackage{subcaption} 
\usepackage[export]{adjustbox} 
\usepackage[normalem]{ulem} 
\sectionfont{\sffamily\bfseries\upshape\large}
\subsectionfont{\sffamily\bfseries\upshape\normalsize} 
\subsubsectionfont{\sffamily\mdseries\upshape\normalsize}
\makeatletter
\renewcommand\@seccntformat[1]{\csname the#1\endcsname.\quad}

\newcommand{\ome}{v}
\newcommand{\transf}{\accentset{\ast}}

\newenvironment{nalign}{
    \begin{equation}
    \begin{aligned}
}{
    \end{aligned}
    \end{equation}
    \ignorespacesafterend
}

\newenvironment{nalign*}{
    \begin{equation*}
    \begin{aligned}
}{
    \end{aligned}
    \end{equation*}
    \ignorespacesafterend
}

\usepackage{bm}

\def\app#1#2{%
  \mathrel{%
    \setbox0=\hbox{$#1\sim$}%
    \setbox2=\hbox{%
      \rlap{\hbox{$#1\propto$}}%
      \lower1.1\ht0\box0%
    }%
    \raise0.25\ht2\box2%
  }%
}

\makeatletter
\def\@maketitle{%
  \begin{center}%
  \let \footnote \thanks
    {\large \@title \par}%
    {\normalsize
      \begin{tabular}[t]{c}%
        \@author
      \end{tabular}\par}%
    {\small \@date}%
  \end{center}%
}
\makeatother

\title{\bf Implicitly Adaptive Importance Sampling}

\author{ Topi Paananen \and Juho Piironen \and Paul-Christian B\"urkner \and Aki Vehtari }

\begin{document}

\maketitle

\begin{abstract}
\noindent Adaptive importance sampling is a class of techniques for finding
good proposal distributions for importance sampling.
Often the proposal distributions are standard probability distributions whose
parameters are adapted based on the mismatch between the current proposal
and a target distribution.
In this work, we present an implicit adaptive importance sampling method that applies to
complicated distributions which are not available in closed form.
The method iteratively matches the moments of a set of Monte Carlo draws
to weighted moments based on importance weights.
We apply the method to Bayesian leave-one-out cross-validation
and show that it performs better than many existing parametric
adaptive importance sampling methods while being computationally inexpensive.

~
\end{abstract}

\section{Introduction}

Importance sampling is a class of procedures
for computing expectations using draws from a proposal
distribution that is different from the distribution over which the 
expectation was originally defined~\citep{robert2013monte}. A primary field of application for
importance sampling is Bayesian statistics where we commonly sample
from the posterior distribution of a probabilistic model as we are unable 
to obtain the distribution in closed form.
After generating a sample from the posterior distribution, it is commonplace to use it
as a proposal distribution for computing a large number of expectations over closely related distributions
for tasks such as bootstrap and leave-one-out cross-validation~\citep{gelfand1992model,gelfand1996model,peruggia1997variability,epifani2008case,vehtari2017practical,giordano2019swiss}.
However, in the presence of influential observations in the data or target distributions
that are difficult to approximate,
such importance sampling procedures may be inefficient or inaccurate.
In order to avoid explicitly generating Monte Carlo draws from each
closely related distribution, it is desirable to find
adaptive importance sampling methods that can utilize the information
in the already generated posterior draws in a computationally efficient manner.

The contributions of this paper can be summarized as follows:
\begin{itemize}
\item We present an implicitly adaptive importance sampling method
for improving the accuracy of a variety of different Monte Carlo approximations of integrals.
The method modifies the proposal distribution implicitly by
iteratively matching the moments of a set of Monte Carlo draws to its weighted moments
based on existing theoretical knowledge of the optimal proposal distributions of different
Monte Carlo estimators.
\item We propose specific adaptations and estimators for
simple Monte Carlo sampling as well as standard and self-normalized importance sampling.
We show that our proposed double adaptation framework for self-normalized importance sampling significantly
improves the accuracy of existing adaptive importance sampling methods in many settings.
\item We also propose to use an existing importance sampling convergence diagnostic
as an acceptance and stopping criterion for the adaptive method, and discuss its applicability
also outside of importance sampling.
\end{itemize}

The proposed method does not require any tuning from the user and is easily automatized and applied
to a variety of different problems.
Because it can be used with arbitrary proposal distributions, it is
most beneficial for complex distributions which are not necessarily of any parametric form.
We demonstrate its usefulness with
Bayesian leave-one-out cross-validation (LOO-CV) and the probabilistic programming framework Stan~\citep{carpenter2017stan}.

\begin{figure}[t]
\centering
\includegraphics[width=\textwidth]{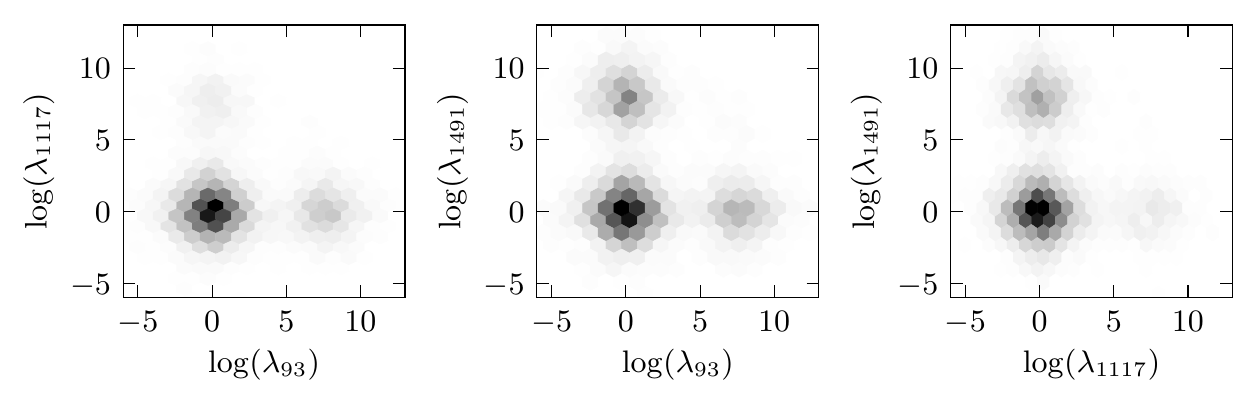}
\caption{Bivariate density plots from the posterior distribution of the logistic regression model
for the Ovarian data. The posterior is 3075-dimensional and is highly multimodal.} \label{fig:ovarian_hex}
\end{figure}

As an illustrative example, we show a Bayesian model posterior that is used in the experiments in Section~\ref{sec:ovarian}. Figure~\ref{fig:ovarian_hex} represents bivariate density plots of the marginal distributions
of three pairs of parameters in the full data posterior distribution. In total, the model has 3075 parameters.
We can use the Monte Carlo sample from the full data posterior
as an importance sampling proposal distribution for computing
cross-validation scores over posterior distributions where single observations have been left out, and our proposed adaptive method
for improving accuracy with a small additional computational cost.
Because the posterior is high-dimensional and
multimodal, standard adaptive methods
that use parametric proposal distributions may require multiple proposal distributions to be efficient,
which is more computationally costly and further complicates the choice of appropriate proposal distributions.
The $\lambda$ parameters in Figure~\ref{fig:ovarian_hex}  are local shrinkage parameters of a logistic regression model with a regularized horseshoe prior on the regression coefficients~\citep{piironen2017rhs}. The parameters are constrained to be positive, so they are sampled in logarithmic space with dynamic Hamiltonian Monte Carlo~\citep{hoffman2014no,Betancourt2017}.
More details are given in Section~\ref{sec:ovarian}.

\subsection{Overview of Importance Sampling} \label{sec:is}

Let us consider an inference problem where a vector of unknown parameters has a probability density
function $p (\boldsymbol{\theta})$. Our task is to estimate integrals of the form
\begin{nalign} \label{eq:integ}
\mu = \mathbb{E}_p [h(\boldsymbol{\theta})] =  \int h(\boldsymbol{\theta}) p (\boldsymbol{\theta}) \mathrm{d} \boldsymbol{\theta} ,
\end{nalign}
where
$h (\boldsymbol{\theta})$ is some function of the parameters $\boldsymbol{\theta}$
that is integrable with respect to $p (\boldsymbol{\theta})$.
These kinds of integrals are ubiquitous in Bayesian inference,
where quantities of interest are computed as expectations over the inferred
posterior distribution of the model.
However, the same formulation is used for many other problems, such as
rare event estimation~\citep{rubino2009rare}, optimal control~\citep{kappen2016adaptive}, and
signal processing~\citep{bugallo2015adaptive}.
Using a set of independent draws $\{ \boldsymbol{\theta}^{(s)} \}_{s = 1}^S$ from $p (\boldsymbol{\theta})$, the \emph{simple Monte Carlo}
estimator of $\mu$ is
\begin{nalign*}
\hat{\mu}_{\text{MC}} = \frac{1}{S} \sum_{s = 1}^S h (\boldsymbol{\theta}^{(s)}) \, , \, \text{when} \,  \, \boldsymbol{\theta}^{(s)} \sim p (\boldsymbol{\theta}) .
\end{nalign*}
In this work, we use the term \emph{draw} to represent a single $\boldsymbol{\theta}^{(s)}$, and
the term \emph{sample} to represent a set of draws $\{ \boldsymbol{\theta}^{(s)} \}_{s = 1}^S$.
If the expectation $\mu$ exists, the simple Monte Carlo estimator is a consistent and unbiased
estimator of $\mu$, meaning
that asymptotically it will converge towards $\mu$ by the strong law of large numbers.
If also the expectation of $h^2$ is finite, the central limit theorem holds, and the asymptotic convergence rate of the simple Monte Carlo
estimator is proportional to $\mathcal{O} (S^{-1/2})$.
Given finite variance, a similar convergence rate holds for uniformly ergodic
Markov chains~\citep[e.g.,][]{roberts2004general}.

In some cases, it is not possible or it is expensive to generate draws from $p (\boldsymbol{\theta})$, but
the expectation $\mu$ is still of interest. 
In this case,
we may generate a sample from a proposal distribution $g (\boldsymbol{\theta})$ and compute the expectation of equation~(\ref{eq:integ}) using the standard importance sampling estimator
\begin{nalign} \label{eq:regIS}
\mathbb{E}_p [h(\boldsymbol{\theta})] \approx \hat{\mu}_{\text{IS}} =  \frac{1}{S} \sum_{s = 1}^S \frac{ p (\boldsymbol{\theta}^{(s)} )}{g (\boldsymbol{\theta}^{(s)})} h (\boldsymbol{\theta}^{(s)})  = \frac{1}{S} \sum_{s = 1}^S w^{(s)} h (\boldsymbol{\theta}^{(s)})  \, , \, \text{when} \,  \, \boldsymbol{\theta}^{(s)}  \sim  g (\boldsymbol{\theta}) .
\end{nalign}
Here, $w^{(s)}$ are called importance weights or importance ratios, and they measure the mismatch between
$p (\boldsymbol{\theta})$ and $g (\boldsymbol{\theta})$ for a specific draw $\boldsymbol{\theta}^{(s)}$.
In principle, the proposal distribution can be any probability distribution which
has the same support as the target distribution $p (\boldsymbol{\theta})$
and is positive whenever $p (\boldsymbol{\theta}) h (\boldsymbol{\theta}) \neq 0$.
The standard importance sampling estimator $\hat{\mu}_{\text{IS}}$ is also 
a consistent and unbiased
estimator of $\mu$ as long as the expectation $\mu$ exists.
Its variance depends largely on the choice of the proposal distribution $g (\boldsymbol{\theta})$. For a good choice, the variance can be smaller than the variance 
of the simple Monte Carlo estimator, but it can also be much larger, or infinite, if the 
choice is less ideal.
In the context of this paper, we will consider the simple Monte Carlo estimator
simply as a special case
of standard importance sampling, where the proposal distribution is $p (\boldsymbol{\theta})$,
the distribution over which the expectation is defined.

A commonly used alternative estimator is the self-normalized importance sampling (SNIS) estimator
\begin{nalign} \label{eq:SNIS}
\hat{\mu}_{\text{SNIS}} =  \frac{ \sum_{s = 1}^S \frac{ p (\boldsymbol{\theta}^{(s)} )}{g (\boldsymbol{\theta}^{(s)})} h (\boldsymbol{\theta}^{(s)})}{\sum_{s = 1}^S \frac{ p (\boldsymbol{\theta}^{(s)} )}{g (\boldsymbol{\theta}^{(s)})}} = \frac{ \sum_{s = 1}^S w^{(s)} h (\boldsymbol{\theta}^{(s)})}{\sum_{s = 1}^S w^{(s)}}  \, , \, \text{when} \,  \, \boldsymbol{\theta}^{(s)} \sim g (\boldsymbol{\theta}) .
\end{nalign}
This estimator is more generally applicable, because it can be used even if the normalization constants
of the densities $p(\boldsymbol{\theta})$ or $g(\boldsymbol{\theta})$ are not known.
The self-normalized estimator is also consistent, but has a small bias of order $\mathcal{O} (1/S)$~\citep{mcbook}.
All three introduced Monte Carlo estimators are consistent, and thus
converge to the true value $\mu$ asymptotically as $S \rightarrow \infty$, if $\mu$ itself exists. However, there are many cases where these estimators can have poor pre-asymptotic behaviour despite having
asymptotically guaranteed convergence~\citep{vehtari2015pareto}. That is, in cases with poor pre-asymptotic behaviour, convergence for any achievable finite set of $S$ draws may be so bad, that we cannot get sufficiently accurate results in reasonable time. We discuss this issue in Section~\ref{sec:diagnostics}.

To unify the notation and nomenclature of the different Monte Carlo estimators, we define
the ratio of the target density $p (\boldsymbol{\theta} )$ and the proposal density $g (\boldsymbol{\theta} )$
as the \emph{common}
importance weights because they do not depend on the function $h (\boldsymbol{\theta} )$
whose expectation is computed:
\begin{nalign} \label{eq:commonw}
w = w(\boldsymbol{\theta}) =  \frac{p(\boldsymbol{\theta})}{g(\boldsymbol{\theta})}  .
\end{nalign}
Analogously, we define the product
\begin{nalign} \label{eq:intw}
\ome = \ome(\boldsymbol{\theta}) = \frac{p(\boldsymbol{\theta})}{g(\boldsymbol{\theta})} h(\boldsymbol{\theta})
\end{nalign}
as the \emph{expectation-specific} importance weights for the expectation $\mathbb{E}_p [h(\boldsymbol{\theta})]$.
With this notation, both the simple Monte Carlo and standard importance sampling estimators
are defined as the sample mean of the expectation-specific weights $\ome$.
On the other hand,
the self-normalized importance sampling
estimator in equation~(\ref{eq:SNIS}) is defined as the ratio of the sample means of $\ome $ and $w$.

\subsection{Multiple Importance Sampling}

In this section, we briefly discuss multiple importance sampling, which
forms the basis for many existing adaptive importance sampling techniques~\citep{cornuet2012adaptive,martino2015adaptive,bugallo2017adaptive}.
Multiple importance sampling refers to the case of sampling independently from many proposal distributions~\citep{hesterberg1995weighted, veach1995optimally, owen2000safe}.
Let us denote the $J$ proposal distributions as $\{ g_1, \ldots , g_J \}$
and the number of draws from each as $\{ S_1 , \ldots S_J \}$ such that
$\sum_{j = 1}^J S_j = S$.
The multiple importance sampling estimator is a weighted combination
of the individual importance sampling estimators:
\begin{nalign*}
\hat{\mu}_{\text{MIS}} = \sum_{j = 1}^J \frac{1}{S_j} \sum_{s = 1}^{S_j} \beta_j (\boldsymbol{\theta}^{(j, s)}) \frac{h (\boldsymbol{\theta}^{(j, s)}) p(\boldsymbol{\theta}^{(j, s)})}{g_j (\boldsymbol{\theta}^{(j, s)})} \, , \, \text{when} \,  \, \boldsymbol{\theta}^{(j,s)} \sim g_j (\boldsymbol{\theta}) .
\end{nalign*}
where $\{ \beta_j \}_{j = 1}^J$ is a partition of unity, i.e. for every $\boldsymbol{\theta}$,
$\beta_j (\boldsymbol{\theta}) \geq 0$ and
$\sum_{j = 1}^J \beta_j (\boldsymbol{\theta}) = 1$.
With different ways of choosing the weighting functions $\beta_j$, one can
vary between locally emphasizing one of the proposal distribution $g_j$, or
considering them in a balanced way for every value of $\boldsymbol{\theta}$.

The weighting functions are commonly chosen using a balance heuristic
\begin{nalign*}
\beta_j (\boldsymbol{\theta}) = \frac{S_j g_j (\boldsymbol{\theta})}{\sum_{k = 1}^{J} S_k g_k (\boldsymbol{\theta})} ,
\end{nalign*}
whose variance is proven to be
smaller than the variance of any weighting scheme plus a term that goes to zero as
the smallest $S_j \rightarrow \infty$~\citep{veach1995optimally}.
The balance heuristic is also a quite natural way of combining the draws from different
proposal distributions, as
the importance weights for all draws are computed as if they were sampled
from the same mixture distribution $g_{\alpha} (\boldsymbol{\theta})$
\begin{nalign} \label{eq:dm-mis}
w_{\text{DM-MIS}}^{(j,s)} = \frac{ p (\boldsymbol{\theta}^{(j,s)} )}{g_{\alpha} (\boldsymbol{\theta}^{(j,s)})} = \frac{ p (\boldsymbol{\theta}^{(j,s)} )}{\sum_{j = 1}^J \alpha_j g_j (\boldsymbol{\theta}^{(j,s)})} , \, \, \,  \text{with} \, \, \, \, \alpha_j = \frac{S_j}{S} .
\end{nalign}
With these weights,
the multiple importance sampling estimator is then computed using the usual equations of
standard~(eq.~(\ref{eq:regIS})) or self-normalized~(eq.~(\ref{eq:SNIS}))
importance sampling.

Weights computed using equation~(\ref{eq:dm-mis}) are sometimes called deterministic mixture weights, whereas
an alternative is to only evaluate a single proposal distribution in the denominator:
\begin{nalign*}
w_{\text{s-MIS}}^{(j,s)} = \frac{ p (\boldsymbol{\theta}^{(j,s)} )}{ g_j (\boldsymbol{\theta}^{(j,s)})} \, , \, \text{when} \,  \, \boldsymbol{\theta}^{(j,s)} \sim g_j (\boldsymbol{\theta}) .
\end{nalign*}
Deterministic mixture weighting requires more evaluations of the proposal densities, but the variance
of the resulting estimator is lower~\citep{elvira2019generalized}
There are techniques for improving the efficiency of the balance heuristic~\citep{havran2014optimal,elvira2015efficient,elvira2016heretical,sbert2016variance,sbert2017adaptive,sbert2019generalizing}.
In this work, we use the balance heuristic because of its simplicity and empirically shown good performance.

\subsection{Adaptive Importance Sampling}

Adaptive importance sampling is a general term that refers to an iterative
process for updating a single or multiple proposal distributions to approximate a given
target distribution.
The details of the adaptation can vary in multiple ways, but most
methods consist of three steps: i) generating draws from the proposal distribution(s),
ii) computing the importance weights of the draws, and iii) adapting the proposal distribution(s).

Adaptive importance sampling methods can be categorized in multiple ways, for example
based on the type and number of proposal distributions, weighting scheme, and adaptation
strategy. Most methods use one or more parametric proposal distributions, such as
Gaussian or Student-$t$ distributions. Typical adaptation strategies are
resampling or moment estimation based on the importance weights.
A good review of many different methods and their
classification is presented in~\citet{bugallo2017adaptive}.
For discussion about the convergence of adaptive importance sampling methods, see, for example, \citet{feng2018uniform} and \citet{akyildiz2019convergence}.

Some notable recent algorithms are adaptive multiple importance sampling \citep[AMIS; ][]{cornuet2012adaptive}
and adaptive population importance sampling \citep[APIS; ][]{martino2015adaptive}, which both use
multiple proposal distributions, and weighting based on deterministic mixture weights.
For the adaptation, they rely on weighted moment estimation to adapt
the mean (and possibly covariance) of the proposal distributions.
Population Monte Carlo algorithms are another class of adaptive importance sampling methods, which
typically use weighted resampling as the means of adaptation~\citep{cappe2004population,cappe2008adaptive,elvira2017improving}.

\section{Importance Weighted Moment Matching} \label{sec:trans}

In this section, we present our proposed implicit adaptive importance sampling method,
importance weighted moment matching (IWMM).
We start from the assumption that we have a Monte Carlo sample and
we are computing an expectation of some function as in equation~(\ref{eq:integ}). The sample can be from an arbitrary importance sampling
proposal distribution, or it can be from the actual distribution over which the expectation is defined.
As with any adaptive importance sampling method, our motivation is that the accuracy of the expectation using the
current sample is not good enough.
The situation where the proposed framework is most beneficial is when
the sample is from a \emph{relatively good}, complex proposal distribution with no closed form and which is expensive to sample from.
Situations like this arise often in Bayesian inference, when a Monte Carlo sample from the
full data posterior distribution has been sampled, and model
evaluations using cross-validation or bootstrap are of interest~\citep{gelfand1992model,gelfand1996model,peruggia1997variability,epifani2008case,vehtari2017practical,giordano2019swiss}.
In this case, implicit adaptation of the proposal distribution can benefit from the existing sample and
improve Monte Carlo accuracy with a small computational cost.

There are similar approaches that adapt proposal distributions nonparametrically using e.g. kernel density estimates~\citep{zhang1996nonparametric}. With the implicit adaptation, we avoid both the resampling and density estimation steps.
Unbiased path sampling by~\citet{rischard2018unbiased} can also use arbitrary proposal distributions, but their approach
requires a considerable amount of tuning from the user.

\subsection{Target of Adaptation} \label{sec:target}

Let us recap the three general steps of adaptive importance sampling, which are
i) generating draws from the proposal distribution(s),
ii) computing the importance weights of the draws, and iii) adapting the proposal distribution(s)
based on the weights.
In our proposed method, step i) is omitted because we do not resample
during adaptation, and instead use the same sample that is transformed directly.
For this reason, the method can be used with any Monte Carlo sample whose
probability density function is known.

For step ii),
unlike most adaptive importance sampling methods, we are not primarily interested in
perfectly adapting the proposal distribution to the distribution over which the expectation is defined, which
is often called the \emph{target distribution} in the importance sampling literature.
While this is a reasonable goal in many cases, in sections~\ref{sec:toyexperim}~and~\ref{sec:poisson} we show examples
where sampling from the target distribution itself
leads to extremely biased estimates.
Instead, we are mainly interested in adapting to the theoretical optimal proposal distribution
of a given expectation $\mathbb{E}_p [h(\boldsymbol{\theta})]$, which
depends on three things: the distribution $p (\boldsymbol{\theta})$ over which the expectation is defined, the function $h (\boldsymbol{\theta})$ whose
expectation is computed, and the Monte Carlo estimator that is used.
For the standard importance sampling estimator,
the optimal proposal distribution is proportional to~\citep{kahn1953methods}
\begin{nalign} \label{eq:isopt}
g_{\text{IS}}^{\text{opt}} & (\boldsymbol{\theta})  \propto p (\boldsymbol{\theta}) \, |h(\boldsymbol{\theta}) \, | ,
\end{nalign}
and for the self-normalized importance sampling estimator, it is~\citep{hesterberg1988advances}
\begin{nalign} \label{eq:snisopt}
g_{\text{SNIS}}^{\text{opt}} & (\boldsymbol{\theta})  \propto p (\boldsymbol{\theta}) \, |h(\boldsymbol{\theta}) - \mathbb{E}_p [h(\boldsymbol{\theta})]\, | .
\end{nalign}
The more complicated form for the self-normalized estimator is due to the requirement of
accurate estimation of both the numerator and denominator of equation~(\ref{eq:SNIS}) simultaneously.

In order to approach the optimal proposal distribution, we define the importance weights for adaptation as follows:
When using the standard importance sampling (or simple Monte Carlo) estimator, 
we use the absolute values of the \emph{expectation-specific} weights of equation~(\ref{eq:intw})
for adaptation, because they quantify the mismatch
between the current proposal and the optimal proposal density.
For the self-normalized importance sampling estimator,
    we recommend separate adaptations for the numerator and the denominator,
    using the absolute values
    of the expectation-specific weights and the \emph{common} importance weights, respectively.
    The results of the two adaptations
    are combined with multiple importance sampling to approximate the optimal proposal density
    of equation~(\ref{eq:snisopt}).
To combine the two adaptations into an efficient proposal distribution, we use an approximation based on superimposing a simpler distribution on top of the optimal proposal:
\begin{nalign} \label{eq:5050prop}
g_{\text{SNIS}}^{\text{split}} (\boldsymbol{\theta}) \propto | h(\boldsymbol{\theta}) | p (\boldsymbol{\theta})  + \mathbb{E}_p [h(\boldsymbol{\theta})] p (\boldsymbol{\theta}) .
\end{nalign}
We call this the \emph{split proposal} density, because it splits the piecewise defined
density of equation~(\ref{eq:snisopt}) into two clear components.
The first component is proportional to equation~(\ref{eq:isopt}) and is thus approximated
with the adaptation using the absolute expectation-specific weights.
The second component is proportional to $p (\boldsymbol{\theta})$ and is reached with
the adaptation using the common weights.

Equation~(\ref{eq:5050prop}) is a convenient approximation to the optimal proposal
of self-normalized importance sampling
because it has
similar tails while being simpler to sample from
because it has two clear components, whereas the density in
equation~(\ref{eq:snisopt}) can easily be multimodal even when the expectation is defined over a unimodal distribution.
The drawback of this approximation is that it places unnecessary probability mass in areas where
$h(\boldsymbol{\theta}) \approx \mathbb{E}_p [h(\boldsymbol{\theta})] $, thus losing some efficiency.
However, generally the more distinct $p (\boldsymbol{\theta})$ is from
$p (\boldsymbol{\theta}) |h (\boldsymbol{\theta})|$, the smaller
$\mathbb{E}_p [h(\boldsymbol{\theta})]$ becomes and hence the approximation becomes closer to the optimal form.
Fortunately, these are the cases
when adaptive importance sampling techniques are most needed.
In Figure~\ref{fig:optimal_vs_smm_3} in Appendix~\ref{sec:optimality}, we show an example
of this phenomenon.

Because equation~(\ref{eq:5050prop}) is a sum of two terms, it is essentially a multiple
importance sampling proposal distribution with two components.
The number of Monte Carlo draws that should be allocated to each component
depends on the properties of $p(\boldsymbol{\theta})$ and $h(\boldsymbol{\theta})$.
A conservative choice is to allocate the same number to both terms.
When $h(\boldsymbol{\theta})$ is nonnegative, this is actually the optimal allocation, because
both terms in equation~(\ref{eq:5050prop}) integrate to $\mathbb{E}_p [h(\boldsymbol{\theta})]$.
With this allocation, 
multiple sampling with the balance heuristic is safe in the sense that the asymptotic variance of the
estimator is never larger than $2$ times the variance of
standard importance sampling using
the better component~\citep{he2014optimal}.
We note that the double adaptation and combining with equation~(\ref{eq:5050prop}) is possible
and beneficial
also with existing parametric adaptive importance sampling methods.
We demonstrate this in the experiments section.

\subsection{Affine Transformations}

Step iii) of adaptive importance sampling is the adaptation of the current proposal distribution(s).
Two commonly used adaptation techniques are weighted resampling and moment estimation~\citep{bugallo2017adaptive}.
In parametric adaptive importance sampling methods, weighted moment estimation is used
to update the location and scale parameters of the parametric proposal distribution.
We employ a similar idea, but instead directly transform the Monte Carlo sample
using an affine transformation.
This enables adaptation of proposal
distributions which do not have a location-scale parameterisation or even a
closed form representation. The only requirement is that
the (possibly unnormalized) probability density is computable.
This property makes it
useful in many practical situations
where a Monte Carlo sample has been generated with probabilistic programming
tools, or other Markov chain Monte Carlo methods.
By using a reversible transformation,
we can compute the probability density function of the transformed draws
in the adapted proposal with the Jacobian
of the transformation.

In this work, we consider simple affine transformations, because both the
transformation and its Jacobian are computationally cheap.
Consider approximating the expectation $\mathbb{E}_p [h(\boldsymbol{\theta})]$
with a set of draws $\{ \boldsymbol{\theta}^{(s)} \}_{s = 1}^S$
from an arbitrary proposal distribution $g (\boldsymbol{\theta})$ (which can also be $p (\boldsymbol{\theta})$ itself).
For a specific draw $\boldsymbol{\theta}^{(s)}$,
a generic affine transformation includes a square matrix $\mathbf{A}$ representing a linear map, and translation vector $\mathbf{b}$:
\begin{nalign} \label{eq:affinetrans}
T: \boldsymbol{\theta}^{(s)} \mapsto \mathbf{A} \boldsymbol{\theta}^{(s)} + \mathbf{b} =: \transf{\boldsymbol{\theta}}^{(s)}.
\end{nalign}
Because the transformation is affine and the same for all draws, the new implicit density $g_T$ evaluated
at every $\transf{\boldsymbol{\theta}}^{(s)}$ changes by a constant, namely
the inverse of the determinant of the Jacobian, $|\mathbf{J}_T|^{-1} = \left | \frac{\mathrm{d} T (\boldsymbol{\theta})}{\mathrm{d} \boldsymbol{\theta}} \right |^{-1}$.
Note that to compute the inverse, the matrix $\mathbf{A}$ must be invertible.
After the transformation, the implicit probability density of the adapted proposal $g_T$ for the
transformed draw $\transf{\boldsymbol{\theta}}^{(s)}$ is
\begin{nalign*}
g_T(\transf{\boldsymbol{\theta}}^{(s)}) = g(\boldsymbol{\theta}^{(s)}) |\mathbf{J}_T|^{-1} .
\end{nalign*}
If the original proposal density $g$ was known only up to an unknown normalizing constant, the
adapted proposal $g_T$ has that same unknown constant.
This is crucial in order to be able to use the split proposal distribution
of equation~(\ref{eq:5050prop}) for self-normalized importance sampling.

To reduce the mismatch between a Monte Carlo sample $\{ \boldsymbol{\theta}^{(s)} \}_{s = 1}^S$ and a given
adaptation target, we consider three affine moment matching transformations
with varying degrees of simplicity.
The transformation have a similar idea as \emph{warp} transformations used for bridge sampling by~\citet{meng2002warp}.
The importance weights used in the transformations can be either the common weights or the
expectation-specific weights, depending on what the adaptation target is, as discussed in Section~\ref{sec:target}.
We define the first transformation, $T_1$, to match the mean of the sample to its importance
weighted mean:
\begin{nalign*}
 \transf{\boldsymbol{\theta}}^{(s)} & = T_1 ( \boldsymbol{\theta}^{(s)} ) =  \boldsymbol{\theta}^{(s)} - \overline{\boldsymbol{\theta}}  +  \overline{\boldsymbol{\theta}}_{w},  \\
\overline{\boldsymbol{\theta}} & = \frac{1}{S} \sum_{s=1}^S \boldsymbol{\theta}^{(s)}  , \\ \overline{\boldsymbol{\theta}}_{w} & = \frac{\sum_{s=1}^S w^{(s)} \boldsymbol{\theta}^{(s)}}{\sum_{s=1}^S w^{(s)}} .
\end{nalign*}
We define $T_2$ to match the marginal variance in addition to the mean:
\begin{nalign*}
 \transf{\boldsymbol{\theta}}^{(s)} & = T_2 ( \boldsymbol{\theta}^{(s)} ) =  \mathbf{v}_{w}^{1/2} \circ \mathbf{v}^{-1/2} \circ ( \boldsymbol{\theta}^{(s)} - \overline{\boldsymbol{\theta}} ) +  \overline{\boldsymbol{\theta}}_{w}, \\
\mathbf{v}  & = \frac{1}{S} \sum_{s=1}^S ( \boldsymbol{\theta}^{(s)} - \overline{\boldsymbol{\theta}} ) \circ ( \boldsymbol{\theta}^{(s)} - \overline{\boldsymbol{\theta}} ) , \\ 
\mathbf{v}_{w} &  = \frac{\sum_{s=1}^S w^{(s)} ( \boldsymbol{\theta}^{(s)} - \overline{\boldsymbol{\theta}} ) \circ ( \boldsymbol{\theta}^{(s)} - \overline{\boldsymbol{\theta}} ) }{\sum_{s=1}^S w^{(s)}}  ,
\end{nalign*}
where $\circ$ refers to a pointwise product of the elements of two vectors.
The final transformation,
$T_3$, matches the covariance and the mean:
\begin{nalign*}
 \transf{\boldsymbol{\theta}}^{(s)} & = T_3 ( \boldsymbol{\theta}^{(s)} ) =  \mathbf{L}_{w} \mathbf{L}^{-1} ( \boldsymbol{\theta}^{(s)} - \overline{\boldsymbol{\theta}} ) +  \overline{\boldsymbol{\theta}}_{w}, \\
\mathbf{L} \mathbf{L}^{\mathsf{T}} & = \boldsymbol{\Sigma}  = \frac{1}{S} \sum_{s=1}^S ( \boldsymbol{\theta}^{(s)} - \overline{\boldsymbol{\theta}} ) ( \boldsymbol{\theta}^{(s)} - \overline{\boldsymbol{\theta}} )^{\mathsf{T}}  , \\ 
\mathbf{L}_{w} \mathbf{L}_{w}^{\mathsf{T}} & = \boldsymbol{\Sigma}_{w} = \frac{\sum_{s=1}^S w^{(s)} ( \boldsymbol{\theta}^{(s)} - \overline{\boldsymbol{\theta}}_{w} ) ( \boldsymbol{\theta}^{(s)} - \overline{\boldsymbol{\theta}}_{w} )^{\mathsf{T}} }{\sum_{s=1}^S w^{(s)}}  .
\end{nalign*}

If the weights are available with the correct normalization, the weighted moments can be computed using
standard importance sampling, but for a more general case, we show the self-normalized estimators of the weighted moments.
When relying on self-normalized importance sampling, we recommend two separate adaptations,
as discussed in Section~\ref{sec:target}.
We
perform both adaptations separately with the full Monte Carlo sample, but for
the multiple importance sampling estimator of equation~(\ref{eq:5050prop}) we
split the existing sample into two equally sized parts to avoid causing bias from using the same draws twice.

The three affine transformations are defined from simple to complex in terms of the \emph{effective} sample size
required to accurately compute the moments~\citep{kong1992note,martino2017effective,chatterjee2018sample,elvira2018rethinking}. Especially in the third transformation,
the weighted covariance can be impossible to compute if the variance of the weight distribution is large.
For this reason, we first iterate only $T_1$ repeatedly, and move on to $T_2$ and $T_3$
only when $T_1$ is no longer helping.
To determine this, we use finite sample diagnostics which will be discussed next.

\subsection{Stopping Criteria and Diagnostics} \label{sec:diagnostics}

Even if an (adaptive) importance sampling procedure has good asymptotic properties
or the used proposal distribution guarantees finite variance by construction, its
pre-asymptotic behaviour can be poor. Because of this, finite
sample diagnostics are extremely important for assessing pre-asymptotic behaviour.
For example,~\citet{vehtari2015pareto} demonstrate importance sampling cases with asymptotically finite variance, but pre-asymptotic behavior indistinguishable from
cases with unbounded importance weights or infinite variance.
\citet{vehtari2015pareto} propose a finite sample diagnostic based on fitting a
generalized Pareto distribution to the upper tail of the distribution of the importance weights.
Because theoretically the shape parameter $k$ of the generalized Pareto distribution determines the number of its finite moments,
the fitted distribution and its shape parameter $\hat{k}$ are useful for estimating practical pre-asymptotic
convergence rate.
The authors propose $\hat{k} = 0.7$ as an upper limit of practically useful pre-asymptotic convergence. The stability of (self-normalised) importance sampling can be improved by replacing the largest weights with ordered statistics of the generalized Pareto distribution estimated already for the diagnostic purposes.

The Pareto $\hat{k}$ diagnostic can also be used as a stopping criterion for adaptive importance sampling methods
in order to not run the adaptation excessively long and waste computational resources.
In addition to that, in the
importance weighted moment matching method we use the diagnostic for
estimating whether a specific transformation
improves the proposal distribution or not.

We use the Pareto diagnostic as follows.
First, we compute the Pareto $\hat{k}$ diagnostic value
for the original (common or expectation-specific) weights. After a transformation ($T_1$, $T_2$ or $T_3$) we recompute the diagnostic, and
only accept the transformation if the diagnostic value has decreased. If it has, the transformation is accepted and
the weights and diagnostic value are updated.
We begin the adaptation by repeating transformation $T_1$, and
only when it is no longer accepted, we move on to
attempt transformation $T_2$, and eventually $T_3$.
As a criterion for stopping the whole algorithm, we use the diagnostic value $\hat{k} = 0.7$ as recommended by~\citet{vehtari2015pareto}
as a practical upper limit for useful accuracy.

The full importance weighted moment matching algorithm for standard importance sampling or simple Monte Carlo sampling
is presented in Algorithm~\ref{alg:mm-is-general}.
When using self-normalized importance sampling, the algorithm
is very similar, with the exception of
having two separate adaptations and combining them with multiple
importance sampling in the end. It
is presented as Algorithm~\ref{alg:mm-snis-general} in Appendix~\ref{sec:appendix-alg}.

\begin{algorithm*}[htb]
\caption{\em Moment matching for standard importance sampling}\label{alg:mm-is-general}
\begin{algorithmic}[1]
\STATE \textbf{Input:} $k_{\text{threshold}}$, proposal density $g$, draws $\{ \boldsymbol{\theta}_i^{(s)} \}_{s = 1}^S$ from $g$
\STATE Compute expectation-specific weights $\{  \ome^{(s)}  \}_{s = 1}^S$ and compute diagnostic $\hat{k}$;

\WHILE{$\hat{k} > k_{\text{threshold}}$}
\FOR{$j$ in $1:3$}

\STATE Transform the draws with $T_j: \boldsymbol{\theta}^{(s)} \mapsto \transf{\boldsymbol{\theta}}^{(s)}$ using absolute expectation-specific weights;
\STATE Recompute expectation-specific weights $\{  \transf{\ome}^{(s)}  \}_{s = 1}^S$ and $\hat{\transf{k}}$;
\IF{$\hat{\transf{k}} < \hat{k}$}
\STATE Accept the transformation and update $\{ \boldsymbol{\theta}^{(s)} \}_{s = 1}^S = \{ \transf{\boldsymbol{\theta}}^{(s)} \}_{s = 1}^S$, $\{  \ome^{(s)} \}_{s = 1}^S = \{ \transf{\ome}^{(s)} \}_{s = 1}^S$, and $\hat{k} = \hat{\transf{k}}$;
\STATE Exit for loop;
\ELSE
\STATE Discard the transformation;
\ENDIF

\IF{$j == 3$}
\STATE Moment matching failed because $\hat{k} > k_{\text{threshold}}$, end algorithm with a warning about sampling inaccuracy;
\ENDIF

\ENDFOR
\ENDWHILE
 
\STATE Moment matching succeeded, compute expectation $\mathbb{E}_p [h(\boldsymbol{\theta})]$ using equation~(\ref{eq:regIS});
 
\end{algorithmic}
\end{algorithm*}

\subsection{Computational Cost} \label{sec:cost}

The computational cost of several popular adaptive importance sampling methods are compared in
Table~\ref{tab:complexity}.
We show here the methods which also use moment estimation and are thus most similar to the proposed importance
weighted moment matching method. For a more exhaustive comparison, see
the review paper by~\citet{bugallo2017adaptive}.
Because of the implicit adaptation in IWMM, the proposal density needs
to be computed only once at the beginning of the algorithm. Thus, the
computational complexity of IWMM is smaller than even the simplest single-proposal adaptive importance sampling
methods.
It is thus well suited for problems where target evaluations are expensive.
We note that IWMM could also replicate the proposal distribution of consecutive transformations
in a similar fashion as adaptive multiple importance sampling
to increase performance at the cost of increased computational complexity~\citep{cornuet2012adaptive}.
However, this may cause bias because there is no resampling. We leave this as possible
direction for future research.

\begin{table*}[tb]
\centering
\caption{Total computational costs of different adaptive importance sampling algorithms after $T$ iterations. $S$ represents the number of draws sampled per iteration from each proposal distribution, except for IWMM which does not resample. $N$ represents the number of proposal distributions.}
\label{tab:complexity}
\begin{tabular}{ l l l }
\toprule
Algorithm & Target evaluations & Proposal evaluations  \\
\midrule
importance weighted \\ moment matching (IWMM)                         & $\mathcal{O}(ST)$  & $\mathcal{O}(S)$ \\
\hline
single-proposal adaptive \\ importance sampling (AIS)                        & $\mathcal{O}(ST)$  & $\mathcal{O}(ST)$  \\
\hline
adaptive multiple \\ importance sampling (AMIS)                         & $\mathcal{O}(ST)$  & $\mathcal{O}(ST^2)$   \\
\hline
adaptive population \\ importance sampling (APIS)                         & $\mathcal{O}(NST)$  & $\mathcal{O}(N^2 S T)$    \\

\bottomrule
\end{tabular}
\end{table*}

If the importance weights have large variance, the moment matching transformations can be noisy
because of inaccurate computation of the weighted moments.
There are two principal ways to remediate this.
First, increasing the number of draws generally increases the accuracy
of the computed moments.
Second, the importance weights used for computing the weighted moments
can be regularized with truncation or smoothing methods~\citep{ionides2008truncated,koblents2015population,miguez2018analysis,vehtari2015pareto,bugallo2017adaptive}.
In the experiments section, we demonstrate that
the accuracy of the moment matching can be improved with Pareto smoothing from~\citet{vehtari2015pareto}.

Another shortcoming of the method is that the adaptation target is not always well characterized by its first and second moments, and
the target and proposal distributions can differ in several
characteristics, such as tail thickness, correlation structure, or
number of modes.
For complex targets, more
elaborate transformations may be needed
to reach a good enough proposal distribution.
That being said, it is not necessary to match all characteristics of the
proposal distribution to the target for importance sampling to be effective.

\section{Experiments} \label{sec:experiments}

In this section,
the proposed implicit adaptation method is illustrated with a variety of
numerical experiments using leave-one-out cross-validation (LOO-CV) as an example application.
With both simulated and real data sets, we evaluate
the predictive performance of different Bayesian models using
leave-one-out cross-validation, and demonstrate
the improvements that the implicit adaptation methods can provide.
We also compare it to existing adaptive importance sampling methods
that use parametric proposal distributions.

All of the simulations were done in R~\citep{rlang}, and
the models were fitted using \texttt{rstan}, 
the R interface to the Bayesian inference package Stan~\citep{carpenter2017stan,rstan}.
To sample from the posterior of each model, we ran four Markov chains using a dynamic Hamiltonian Monte Carlo (HMC)
algorithm~\citep{hoffman2014no,Betancourt2017} which is the default in Stan.
We monitor convergence of the chains with the split-$\widehat{R}$ potential scale reduction factor from~\citet{Vehtari+etal:2019:Rhat} and by checking for divergence transitions, which is a diagnostic
specific to adaptive HMC.
We note that the finite sample behaviour of Monte Carlo integrals
depends on the algorithm used to generate the sample.
For example, if one uses an MCMC algorithm less efficient than HMC, the
resulting Monte Carlo approximations will generally be worse than those illustrated in the next sections.
R and Stan codes of the experiments and the used data sets are available on Github (\url{https://github.com/topipa/iter-mm-paper}).

Because probabilistic programming tools generally give only unnormalized posterior densities,
we mostly focus on self-normalized importance sampling.
As the default case, we take the situation that Monte Carlo
draws are available from the full data posterior distribution, and these are adapted using our proposed
method. We note that leave-one-out cross-validation in this setting is a special case
such that the double adaptation which is discussed in Section~\ref{sec:target}
is not needed even when using self-normalized importance sampling.
The split proposal of equation~(\ref{eq:5050prop}) is still used, but
the other term uses the full data posterior draws.
To help the reader in understanding or implementing the methods,
we have presented
the basics of Bayesian leave-one-out cross-validation
as well as instructions for implementing the proposed methods in Appendix~\ref{appendix:loo}.
In addition to importance sampling, we also discuss simple Monte Carlo sampling results when sampling from each
leave-one-out posterior explicitly.

By default, we use Pareto smoothing to stabilize importance weights, but
we also present results without smoothing~\citep{vehtari2017practical,vehtari2015pareto}.
This enables us to also monitor the reliability of the Monte Carlo estimates
using the Pareto $\hat{k}$ diagnostics. We show that the
diagnostics accurately identify convergence problems
in not only importance sampling, but also when using the simple Monte Carlo estimator
or adaptive importance sampling algorithms.
Based on~\citet{vehtari2015pareto}, we use $\hat{k} = 0.7$ as
an upper threshold to indicate practically useful finite sample convergence rate.

We compare our proposed method to several existing adaptive importance sampling methods.
For comparison, we chose algorithms that are conceptually similar to our proposed implicit adaptation method.
As the first comparison, we have generic adaptive importance sampling methods, which
use a single proposal distribution and
adapt the location and scale parameters of this distribution using weighted moment estimation.
As the proposal distribution we have either a multivariate Gaussian distribution, or
a Student-$t_3$ distribution. Moreover, we test these algorithms in the traditional way
of adapting using the common importance weights, and also using our proposed double adaptation, resulting
in 4 different algorithms.
To compare to a more powerful and computationally expensive algorithm, we chose
adaptive multiple importance sampling~\citep[AMIS;][]{cornuet2012adaptive}, which
uses
multiple proposal distributions and deterministic mixture weighting, increasing
the number of proposal distributions over time.
Also for this algorithm, we test 4 versions by having either Gaussian or Student-$t_3$ distributions
as well as with and without double adaptation.
We start all the parametric adaptive methods with mean and covariance
estimated from a sample from the full data posterior.
Also for the parametric adaptive importance sampling methods,
we use the Pareto $\hat{k}$ diagnostic to determine when to stop the algorithm.
For all eight algorithms, we adapt both the mean and covariance if it is feasible, but
for very high-dimensional distributions we only adapt the mean, because
otherwise the adaptation is unstable given the used sample sizes.

Sections~\ref{sec:toyexperim} and \ref{sec:poisson} show low-dimensional examples where
the function $h$ whose expectation is being computed gets large values in the tails
of the distribution over which the expectation is being computed. These cases highlight the importance
of our proposed double adaptation, as the target densities are available in unnormalized form.
Sections~\ref{sec:linreg}~and~\ref{sec:ovarian} show correlated and high-dimensional examples which are significantly
more difficult.
In Section~\ref{sec:ovarian}, the
distribution over which the expectation is defined is also multimodal.
In these cases, we demonstrate the usefulness of using a complex non-parametric proposal distribution
instead of Gaussian or Student-$t$ densities.

\subsection{Experiment 1: Gaussian Data with a Single Outlier} \label{sec:toyexperim}

In this section, we demonstrate with a simple example what happens when
we try to assess the predictive performance of a misspecified model.
We emphasize that even though this is a simple example, it still provides valuable
insight for real world data and models as
evaluating misspecified models is an integral part of any Bayesian modelling process.
In terms of Monte Carlo sampling, this is an example of an expectation~(\ref{eq:integ}) where
the largest values of the function $h$ are in the tails of the target distribution $p$.

We generate 29 observations from a standard normal distribution, and manually
set the value for a 30'th observation to introduce an outlier.
This mimics a situation where the true data generating mechanism has thicker tails
than the assumed observation model.
Keeping the randomly generated observations fixed, we repeat the experiment
for different values of the outlier ranging from $y_{30} = 0$ to $y_{30} = 20$.
We
model the data with a Gaussian distribution with unknown mean and variance, generate
draws from the model posterior, and evaluate the predictive ability of the model using
leave-one-out cross-validation.

For all 30 observations, represented jointly by the vector $\mathbf{y}$, 
the model is thus
\begin{nalign*}
\mathbf{y} \sim \text{Normal} (\mu,\sigma^2)
\end{nalign*}
with mean $\mu$ and standard deviation $\sigma$. 
We set improper uniform priors on $\mu$ and $\log(\sigma)$.
In this model, the posterior predictive distribution $p (\widetilde{y} \mid \mathbf{y})$ is known analytically, and is a
Student $t$-distribution with $n-1$ degrees of freedom, mean at the mean of the data, and
scale $\sqrt{1 + 1/n}$ times the standard deviation of the data, where $n$ is the number of
observations.
Thus, we can compute the Bayesian LOO-CV estimate for the single
left out point analytically via
\begin{nalign*}
\text{elpd}_{\text{loo},i} = \log p (\widetilde{y} = y_i \mid \mathbf{y}_{-i}).
\end{nalign*}

\begin{figure}[t]
\centering
\includegraphics[width=\textwidth]{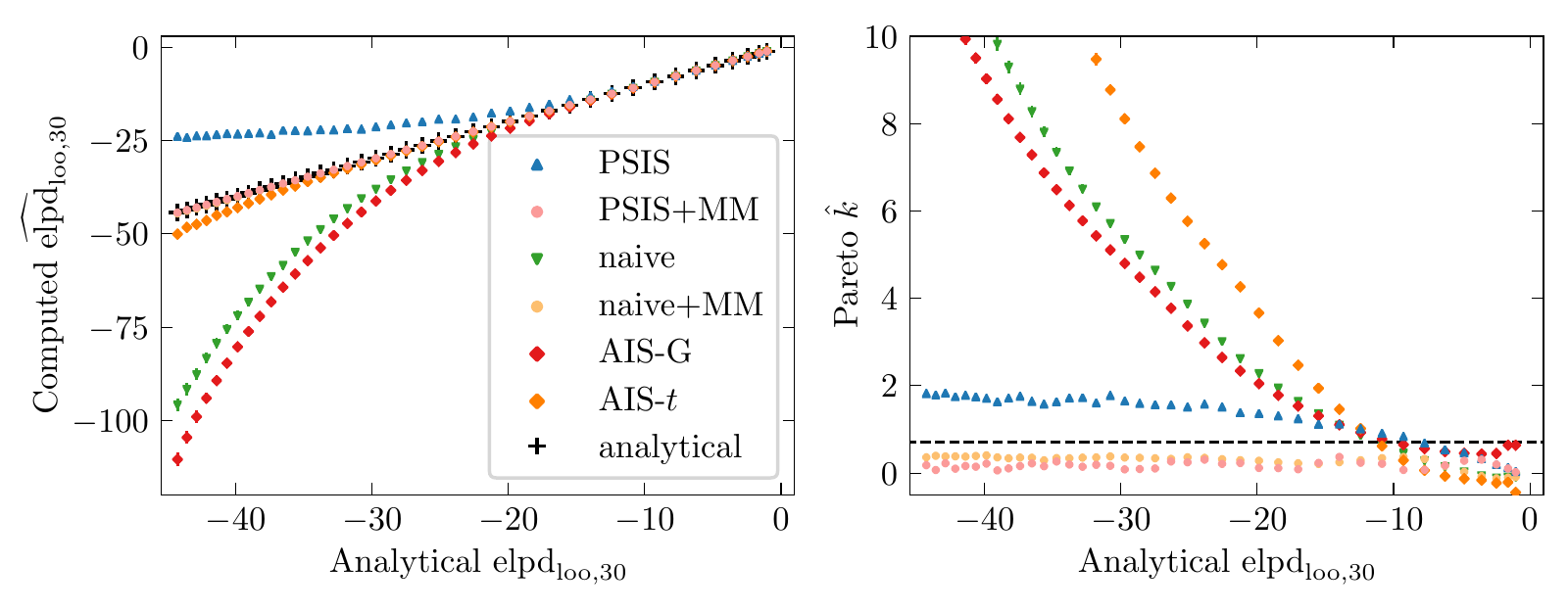}
\caption{Computed $\widehat{\text{elpd}}_{\text{loo},30}$ estimates of the left out observation $y_{30}$ for the normal model
for different values between $y_{30} = 0$ and $y_{30} = 20$.
The black crosses depict the analytical results. The sampling results are averaged from 100 independent Stan runs, and the error bars represent $95 \%$
intervals of the mean across these runs. The dashed line
at $\hat{k} = 0.7$ presents the diagnostic threshold indicating practically useful finite sample convergence rate.} \label{fig:toynormal}
\end{figure}

The left plot of Figure~\ref{fig:toynormal} shows the computed $\widehat{\text{elpd}}_{\text{loo},30}$ estimates
for the $30$'th observation based on different sampling methods, which are
compared to the analytical $\text{elpd}_{\text{loo},30}$ values when the outlier
value is varied.
When the outlier becomes more and more different from the rest of the observations and
the analytical $\text{elpd}_{\text{loo},30}$ decreases, both
the simple Monte Carlo estimate from the true leave-one-out
posterior and the PSIS estimate from the full data posterior
become more and more biased
due to insufficient accuracy in the tails of the posterior predictive distribution.
The same happens to adaptive importance sampling using a single Gaussian proposal (AIS-G),
and to a smaller extent when using a
Student-$t_3$ proposal (AIS-$t$).
Our proposed importance weighted moment matching from either the full posterior (PSIS+MM) or the leave-one-out posterior (naive+MM)
almost perfectly align with the analytical solution.
Also the AIS-G and AIS-$t$ give very accurate results when using our proposed
double adaptation. Similarly, the results of all 4 AMIS algorithms align well with the analytical solution and are omitted in Figure~\ref{fig:toynormal} for improved readability.
While not shown in the plot, also PSIS+MM gives highly biased results if omitting the split proposal
of equation~(\ref{eq:5050prop}).
In Appendix~\ref{appendix:results}, we show the results of a similar experiment, where
the randomly generated points $y_{1}$ to $y_{29}$ are re-generated at every repetition
to show that the results are not just specific to this particular data realization.

The right plot of Figure~\ref{fig:toynormal} shows the Pareto $\hat{k}$ diagnostic values
corresponding to the different algorithms. The diagnostic values are computed from both common
and expectation-specific weights, and the larger is reported. The plot shows that both moment
matching algorithms have $\hat{k} < 0.7$ which indicates good finite sample accuracy.
For all of the other algorithms, the diagnostic value grows over 0.7 when
the problem becomes more difficult, which correlates well with the biased results in the left plot.
From the AIS algorithms, the Student-$t_3$ proposal distribution has much smaller bias
compared to the Gaussian proposal due to its much thicker tails.
Still, the Pareto $\hat{k}$ diagnostic indicates poor finite sample convergence.
When looking at the importance weights of the individual runs, it is indeed clear
that the result is based on only a few Monte Carlo draws from the thick tails of the Student-$t_3$ distribution.
Because of that, the variance between different runs is
large. In the most difficult case when $y_{30} = 20$ and
$\text{elpd}_{\text{loo},30} = -44.3$, the variance of the estimated $\text{elpd}_{\text{loo},30}$
for $\text{AIS-}t$ is more than 1000 times higher than for PSIS+MM.

Figure~\ref{fig:toynormal} highlights the importance of our proposed double adaptation
when some densities are available in unnormalized form.
All of the proposal distributions that we compared
fail without the double adaptation and split proposal of equation~(\ref{eq:5050prop}).
The Student-$t$ proposal does quite well, but
it has high variance because of relying only on a few draws from the tails.
In more high-dimensional situations, it will also fail quicker, as we show later.
AMIS gives good results
even without the double adaptation because it was started from an initial distribution
based on the mean and covariance of the full posterior, and it retains all earlier proposal distributions.
Because this is already close to the target of the second adaptation, it is not needed for the AMIS algorithms
in this low-dimensional example.

\subsection{Experiment 2: Poisson Regression with Outliers} \label{sec:poisson}

In the second experiment, we illustrate with a real data set
how poor finite sample convergence can cause significant errors when estimating predictive performance of models.
The data are from~\citet{gelman2006data}, where the authors describe an experiment that was performed to assess
how efficiently a pest management system reduces the amount of roaches. The target variable $y$ describes the number of roaches
caught in a set of traps in each apartment. The model includes an intercept plus three regression predictors: the number of roaches before treatment, an indicator variable for the treatment or control group, and an indicator variable for whether the building is restricted
to elderly residents.
We will fit a Poisson regression model with a log-link to the data set.
The traps were held in the apartments for different periods of time, so the measurement time is included by adding its logarithm as an offset to the linear predictor.
The model has only 4 parameters, so this is again a quite simple example.

\begin{figure}[t]
\centering
\includegraphics[width=\textwidth]{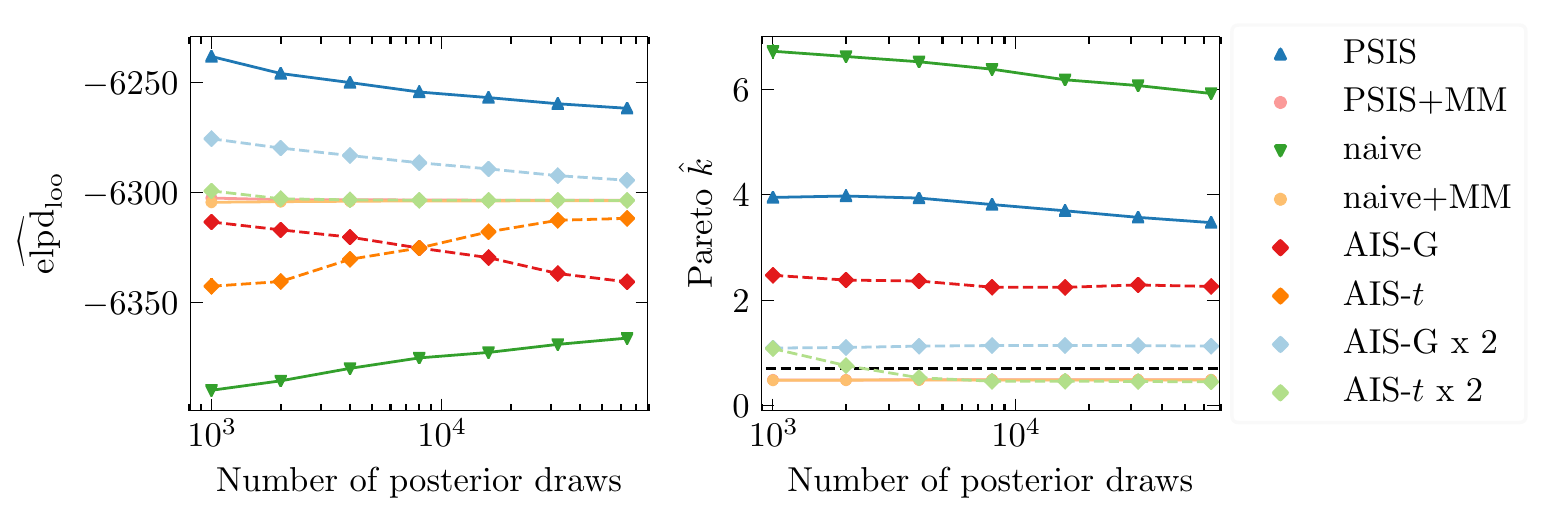}
\caption{Left: Computed $\widehat{\mathrm{elpd}}_{\mathrm{loo}}$
estimates over the whole Roach data set as a function of the number of posterior draws $S$. Right: The mean of the largest Pareto $\hat{k}$ diagnostic values among the observations.
The results are averaged from 100 independent Stan runs, and the error bars represent $95 \%$
intervals of the mean across these runs.
The dashed line
at $\hat{k} = 0.7$ presents the diagnostic threshold indicating practically useful finite sample convergence rate.
The largest Pareto $\hat{k}$ for AIS-$t$ is around 10, and it is left out of the right plot for clarity.} \label{fig:roach_mm_ais}
\end{figure}

On the left side of Figure~\ref{fig:roach_mm_ais} we show the computed $\widehat{\mathrm{elpd}}_{\mathrm{loo}}$
estimates averaged from 100 independent Stan runs as a function of the number of posterior draws $S$.
On the right side, the mean of the largest Pareto $\hat{k}$ diagnostic values out of all of the observations are presented.
The diagnostic is always computed from both
the common and expectation-specific weights, and the larger is reported.
There is a large difference between the PSIS and naive estimates, and they approach each other
very slowly
when increasing $S$, which is due to the poor convergence rate, as indicated by the high Pareto $\hat{k}$ values
on the right side plot.
Importance weighted moment matching from either the full posterior or
leave-one-out posteriors gives reliable estimates with very small error already from 1000 draws.
The accuracy is confirmed by the changed Pareto $\hat{k}$ values which are always below 0.7.
For the single-proposal parametric methods, using the Student-$t_3$ proposal distributions and doing our proposed double adaptation (AIS-$t$~$\times 2$)
gives good results from 2000 draws onwards, but the rest of the methods give highly biased results
even with $S = 64000$.
Contrary to the previous example, now even the double adaptation converges extremely slowly when using a Gaussian
proposal distribution (AIS-G~$\times 2$), which indicates that the posterior distribution is non-Gaussian.
All 4 versions of the AMIS algorithms had Pareto $\hat{k}$ values below 0.7 already with 1000 draws, and
had elpd estimates almost indistinguishable from the importance
weighted moment matching results. These are omitted from~Figure~\ref{fig:roach_mm_ais} for improved readability.

\subsection{Experiment 3: Linear Regression with Correlated Predictor Variables} \label{sec:linreg}

In the previous examples, the used models were quite simple and had a small 
number of parameters.
In this and the following sections, we study the limitations of the importance weighted
moment matching method by considering models with more parameters and
correlated or non-Gaussian posteriors.
The two previous experiments showed that the Pareto $\hat{k}$ diagnostic
is a reliable indicator of finite sample accuracy for adaptive importance sampling methods.
To demonstrate the performance and computational cost of the different adaptive
algorithms, we report the number of leave-one-out (LOO) folds
where the algorithms fail to decrease the $\hat{k}$ diagnostic value
below $0.7$.
In order to get reliable results for these failed LOO folds, the user should generate new MCMC
draws from the LOO posterior, which can be very costly.
We fit all models to the full data set, and report
the number of leave-one-out folds where
the $\hat{k}$ diagnostic value is above $0.7$ when using
the full data posterior directly as a proposal distribution.
These are reported in the column PSIS in Table~\ref{tab:results}.
We run the moment matching algorithm for all these LOO folds, and report
how many $\hat{k}$ values are still above $0.7$ (PSIS+MM).
Similarly, we run the 8 parametric adaptive methods for the same LOO folds.
In Table~\ref{tab:results}, we only show the best performing parametric methods, which are
AMIS with double adaptation using either Gaussian or Student-$t_3$ proposals
(AMIS $\times 2$ and AMIS-$t$ $\times 2$).
In the lower part of Table~\ref{tab:results}, we report run times in seconds for all the reported
algorithms. The run times are based on single core runs with an Intel Xeon X5650 2.67 GHz processor.

For this experiment,
we simulated data from a linear regression model. 
The data consists of $n = 60$ observations of one outcome variable 
and $30$ predictors that are correlated with each other by correlation coefficient
of $\rho = 0.8$. Three of the true regression coefficients are nonzero, and the rest are all zero. Independent Gaussian noise was added to the outcomes $\mathbf{y}$.
Because the predictors are strongly correlated, importance sampling leave-one-out cross-validation is
difficult and we get multiple high Pareto $\hat{k}$ values when using
the full data posterior as the proposal distribution.
The results of Table~\ref{tab:results} show that
already with 2000 posterior draws, the moment matching algorithm
is able to decrease the Pareto $\hat{k}$ values of all LOO folds below $0.7$.
In contrast, none of the parametric algorithms ever succeed in reducing $\hat{k}$ values below $0.7$,
even when increasing the number of draws to 8000.
This highlights the difficulty of adapting to a highly correlated distribution.
Because the moment matching starts from the full data posterior sample,
which is similarly correlated, the moment matching can successfully improve the proposal distribution with a small cost.
The AMIS algorithms were run for 10 iterations to limit the computational cost.
By increasing the number of iterations, they should succeed eventually,
but at a high computational cost.
In Table~\ref{tab:results_raw} in Appendix~\ref{appendix:results}, we show results for importance weighted moment
matching without Pareto smoothing the importance weights. The results are slightly worse compared
to the Pareto smoothing case.

\subsection{Experiment 4: Binary Classification in a Small $n$ Large $p$ Data Set} \label{sec:ovarian}

In the fourth experiment, we have a real microarray Ovarian cancer classification data set with a large number of
predictors and small number of observations.
The data set has been used as a benchmark by several authors~\citep[e.g.][and references]{schummer1999comparative,hernandez2010expectation}.
The data consists of $54$ measurements and has $1536$ predictor variables.
We will fit a logistic regression model
using a regularized horseshoe prior~\citep{piironen2017rhs} on the regression coefficients because
we expect many of them to be zero.
This data set and model are difficult for several reasons.
First, because
the amount of observations is quite low, leaving out single observations
changes the posterior significantly, indicated by a
large number of high Pareto $\hat{k}$ values. Second, because the number of parameters in the model is 3075,
moment matching in the high-dimensional space is difficult.
Third, the posterior distribution of several 
parameters is multimodal, as illustrated in Figure~\ref{fig:ovarian_hex}.
Because of the multimodality, we used Monte Carlo chains of length 1000, and increased the
number of chains when increasing $S$.

When fitting the model to the full data posterior,
Table~\ref{tab:results} shows the number of LOO folds with $\hat{k} > 0.7$
before and after moment matching.
The results show that already with 1000 draws,
PSIS+MM is able to reduce $\hat{k}$ of
many LOO folds below 0.7.
Investing more computational resources
by collecting more posterior draws increases
the moment matching accuracy, and more LOO folds can be improved.
However, even with $8000$ posterior draws some folds have $\hat{k} > 0.7$
after moment matching, and thus the $\widehat{\text{elpd}}_{\text{loo}}$
estimate may not be reliable.
Again, none of the parametric adaptive methods succeed in reducing Pareto $\hat{k}$ values
below 0.7 in 10 iterations. The lower part of Table~\ref{tab:results} also shows
the significantly higher computational time of the AMIS algorithms compared to importance
weighted moment matching.

\begin{table*}[tb]
\centering
\caption{Upper part: Numbers of LOO folds with Pareto $\hat{k}$ diagnostic above 0.7 when the models are fitted to the full data set (lower is better).
Lower part: Average run times in seconds for different algorithms.
Column PSIS corresponds to using the full data posterior directly
as the proposal distribution.
Column PSIS+MM corresponds to importance weighted
moment matching.
Column AMIS $\times 2$
corresponds to adaptive multiple importance sampling with our proposed double adaptation
using Gaussian proposal distributions. Column AMIS-$t$ $\times 2$ is the same, but
using Student-$t_3$ proposals.}
\label{tab:results}
\begin{tabular}{ c }
 \textbf{Folds with} $\hat{k} > 0.7$  \\
\end{tabular}
\begin{tabular}{ l r r r r r }
\toprule
Data and model & Draws & PSIS & PSIS+MM & AMIS $\times 2$  & AMIS-$t$ $\times 2$ \\
\midrule

Section~\ref{sec:poisson}:         & $2000$  & 15.2 & 0 & 0 & 0  \\
\;Roach data                         & $4000$  & 14.7 & 0 & 0 & 0 \\
\;Poisson regression model           & $8000$  & 14.2 & 0 & 0 & 0 \\
\hline                                    
Section~\ref{sec:linreg}:        & $2000$  & 14.0 & 0      & 14.0 & 14.0 \\
\;Correlated predictor variables                              & $4000$  & 13.8 & 0      & 13.8 & 13.8 \\
\;Linear regression model                                 & $8000$  & 13.4 & 0      & 13.4 & 13.4 \\

\hline                                    
Section~\ref{sec:ovarian}:        & $1000$  & 34.8 & 20.1 & 34.8 & 34.8  \\
\;Ovarian cancer data ($n < p$)               & $2000$  & 36.1 & 19.6 & 36.1 & 36.1  \\
\;Logistic regression model         & $4000$  & 34.8 & 16.2 & 34.8 & 34.8 \\
                                  & $8000$  & 34.0 & 11.4 & 34.0 & 34.0  \\
\bottomrule

\end{tabular}

\begin{tabular}{ c }
\, \\
\textbf{Computation times (seconds)}  \\
\end{tabular}

\begin{tabular}{ l r r r r r }
\toprule
Data and model & Draws & PSIS & PSIS+MM &  AMIS $\times 2$  & AMIS-$t$ $\times 2$ \\
\midrule

Section~\ref{sec:poisson}:         & $2000$  & 0 & 39 & 73 & 39  \\
\;Roach data                         & $4000$  & 0 & 70 & 133 & 71 \\
\;Poisson regression model           & $8000$  & 0 & 140 & 330 & 176 \\
\hline     

Section~\ref{sec:linreg}        & $2000$  & 0 & 52      & 196 & 191 \\
\;Correlated predictor variables                              & $4000$  &  0 & 114      & 397 & 382 \\
\;Linear regression model   & $8000$  & 0  & 346      & 932 & 927 \\
   
   \hline
   
Section~\ref{sec:ovarian} & $1000$  & 0 & 199 & 8857 & 11330  \\
\;Ovarian cancer data ($n < p$)             & $2000$  & 0 & 372 & 12289 & 12311  \\
\;Logistic regression model         & $4000$  & 0 & 1558 & 30814 & 27477 \\
      & $8000$  & 0 & 3733 & 54534 & 68082  \\
   
\bottomrule

\end{tabular}
\end{table*}

The used data set and model are complex enough that using naive LOO-CV
by fitting to each LOO fold separately takes
a nontrivial amount of time. Omitting parallelization, the model
fit using Stan took an average of 27 minutes when generating
4000 posterior draws. Naive LOO-CV would be costly as fitting the model 54 times would take around 24 hours.
With the same hardware, standard PSIS took less than a second, but refitting the 34.8 (on average)
problematic LOO folds would take more than 15 hours.
For the problematic LOO folds, the total run time of PSIS+MM was only 26 minutes on average.
This is less time than a single model fit
while decreasing the number of required refits from 34.8 to 16.2 on average, which shows that
the importance weighted moment matching
is computationally efficient.

\section{Conclusion}

We proposed a method for improving the accuracy
of Monte Carlo approximations to integrals via importance sampling and importance weighted moment matching.
By matching the moments of an existing Monte Carlo sample to
its importance weighted moments, the proposal
distribution is implicitly modified and improved.
The method is easy to use and automate for different applications because it has
no parameters that require tuning.
We proposed separate adaptation schemes and estimators for different importance sampling estimators.
In particular, we proposed a novel double adaptation scheme that is beneficial for many existing
adaptive importance sampling methods when relying on the self-normalized importance sampling estimator.

We also showed that the Pareto diagnostic method from~\citet{vehtari2015pareto}
is able to notice poor finite sample convergence for different Monte Carlo estimators
and adaptive algorithms
when taking into account both the common and expectation-specific importance weights.
We also showed that it is useful as a stopping criterion
in adaptive importance sampling methods, reducing computational cost by not running
the algorithm excessively long.

We evaluated the efficacy of the proposed methods in self-normalized importance sampling leave-one-out
cross-validation (LOO-CV), and demonstrated that they can often increase the accuracy
of model assessment and even surpass naive LOO-CV that requires expensive
refitting of the model.
Moreover, in complex or high-dimensional cases we demonstrated that our proposed method has much better performance
compared to existing adaptive importance sampling methods that use
Gaussian or Student-$t_3$ proposal distributions.
Additionally, our method has a small computational cost as it does not require
recomputing proposal densities during iterations.
We also showed that our proposed double adaptation scheme
for self-normalized importance sampling is crucial
for cases where the function whose expectation is being computed has large values
in the tails of the distribution over which the expectation is computed.
We showed that the double adaptation can also significantly improve the performance of
existing parametric adaptive importance sampling methods.

The performance of the proposed implicit adaptation method depends highly
on the goodness of the initial proposal distribution. Bayesian leave-one-out cross-validation or
bootstrap are examples where the full data posterior distribution is
already a good proposal, and moment matching can improve performance
with a small computational cost. In the most complex cases, the simple affine transformation
proposed in this work are not enough to produce a good proposal distribution, and
more complex methods may be required.
Such methods are left for future research.

\section{Acknowledgements}

We thank Michael Riis Andersen, Alejandro Catalina, M{\aa}ns Magnusson and Christian P. Robert for helpful comments and discussions.
We also thank two anonymous reviewers for their helpful suggestions, and
acknowledge the computational resources provided by the Aalto Science-IT project.
We thank Academy of Finland (grants 298742 and 313122),
Finnish Center for Artificial Intelligence,
and Technology Industries of Finland Centennial Foundation (grant 70007503; Artificial Intelligence for Research and Development)
for partial support of this research.

\bibliographystyle{apalike}
\bibliography{iter-mm}

\onecolumn
\clearpage

\begin{appendices}

\section{Moment matching for self-normalized importance sampling} \label{sec:appendix-alg}

\begin{algorithm*}[htb]
\caption{\em Moment matching for self-normalized importance sampling}\label{alg:mm-snis-general}
\begin{algorithmic}[1]
\STATE \textbf{Input:} $k_{\text{threshold}}$, proposal density $g$, draws $\{ \boldsymbol{\theta}_i^{(s)} \}_{s = 1}^S$ from $g$
\STATE Compute common weights $\{  w^{(s)}  \}_{s = 1}^S$ and expectation-specific weights $\{  \ome^{(s)}  \}_{s = 1}^S$, and compute diagnostics $\hat{k}_w$ and $\hat{k}_{\ome}$;

\WHILE{$\hat{k}_{\ome} > k_{\text{threshold}}$}
\FOR{$j$ in $1:3$}

\STATE Transform the draws with $T_j: \boldsymbol{\theta}^{(s)} \mapsto \transf{\boldsymbol{\theta}}^{(s)}$ using absolute expectation-specific weights;
\STATE Recompute expectation-specific weights $\{  \transf{\ome}^{(s)}  \}_{s = 1}^S$ and $\hat{\transf{k}}_{\ome}$;
\IF{$\hat{\transf{k}}_{\ome} < \hat{k}_{\ome}$}
\STATE Accept the transformation and update $\{ \boldsymbol{\theta}^{(s)} \}_{s = 1}^S = \{ \transf{\boldsymbol{\theta}}^{(s)} \}_{s = 1}^S$, $\{  \ome^{(s)} \}_{s = 1}^S = \{ \transf{\ome}^{(s)} \}_{s = 1}^S$, and $\hat{k}_{\ome} = \hat{\transf{k}}_{\ome}$;
\STATE Exit for loop;
\ELSE
\STATE Discard the transformation;
\ENDIF

\IF{$j == 3$}
\STATE Moment matching failed, end algorithm with a warning about sampling inaccuracy;
\ENDIF

\ENDFOR
\ENDWHILE

\WHILE{$\hat{k}_w > k_{\text{threshold}}$}
\FOR{$j$ in $1:3$}

\STATE Transform the draws with $T_j: \boldsymbol{\theta}^{(s)} \mapsto \transf{\boldsymbol{\theta}}^{(s)}$ using common weights;
\STATE Recompute common weights $\{  \transf{w}^{(s)}  \}_{s = 1}^S$ and $\hat{\transf{k}}_w$;
\IF{$\hat{\transf{k}}_w < \hat{k}_w$}
\STATE Accept the transformation and update $\{ \boldsymbol{\theta}^{(s)} \}_{s = 1}^S = \{ \transf{\boldsymbol{\theta}}^{(s)} \}_{s = 1}^S$, $\{  w^{(s)} \}_{s = 1}^S = \{ \transf{w}^{(s)} \}_{s = 1}^S$, and $\hat{k}_w = \hat{\transf{k}}_w$;
\STATE Exit for loop;
\ELSE
\STATE Discard the transformation;
\ENDIF

\IF{$j == 3$}
\STATE Moment matching failed, end algorithm with a warning about sampling inaccuracy;
\ENDIF

\ENDFOR

\ENDWHILE

\STATE Moment matching succeeded, compute common and expectation-specific weights using
the multiple importance sampling density of equation~(\ref{eq:5050prop})
as the proposal density;

\STATE Compute expectation $\mathbb{E}_p [h(\boldsymbol{\theta})]$ using equation~(\ref{eq:SNIS});

\end{algorithmic}
\end{algorithm*}

\section{Bayesian Leave-One-Out Cross-Validation}
\label{appendix:loo}

In this section, we describe importance sampling leave-one-out cross-validation and demonstrate
how the proposed implicit adaptation method can be applied to this problem.

\subsection{Importance Sampling Leave-One-Out Cross-Validation}

After fitting a Bayesian model, it is important to assess its predictive accuracy
as part of the modelling process.
This also enables comparison to other models for model averaging or selection purposes~\citep{geisser1979predictive,hoeting1999bayesian,vehtari2002bayesian,ando2010predictive,vehtari2012survey,Piironen2017}.
Leave-one-out cross-validation (LOO-CV) is a commonly used method for estimating the out-of-sample
predictive ability of a Bayesian model.

As the target measure for the predictive accuracy of a model, we use the
expected log pointwise predictive density (elpd) in a new, unseen data set $\mathbf{\widetilde{y}} = ( \widetilde{y}_1, \ldots , \widetilde{y}_n )$:
\begin{nalign*}
\mathrm{elpd} = \sum_{i = 1}^n \int p_t (\widetilde{y}_i) \log p (\widetilde{y}_i \mid \mathbf{y}) \mathrm{d} \widetilde{y}_i ,
\end{nalign*}
where $p_t (\widetilde{y}_i)$ is the probability distribution of the true data generating mechanism for
the $i$'th observation.
In this paper we use the logarithmic score proposed by~\citet{good1952rational}
as the utility function for evaluating predictive accuracy.
The logarithmic score is a widely used utility function for probabilistic models due to its
suitable theoretical properties~\citep{bernardo1979expected,geisser1979predictive,bernardo1994bayesian,gneiting2007strictly}.

Because we do not know the true data generating mechanism, by making the assumption
that future data has a similar distribution as the measured data, we
can estimate the elpd by means of cross-validation.
LOO-CV is a method for estimating the predictive performance of a model by 
reusing the observations $\mathbf{y} = ( y_1, \ldots , y_n)$ available.
Using the log predictive density as the utility function,
the Bayesian LOO-CV estimator of elpd is
\begin{nalign} \label{eq:naive}
\text{elpd}_{\text{loo}} = & \sum_{i = 1}^{n} \log p (y_i \mid \mathbf{y}_{-i}),
\end{nalign}
where $p (y_i \mid \mathbf{y}_{-i})$ is the LOO posterior predictive density when leaving out the observation~$y_i$:
\begin{nalign} \label{eq:loopd}
p (y_i \mid \mathbf{y}_{-i}) = \int p (y_i \mid \boldsymbol{\theta}) p (\boldsymbol{\theta} \mid \mathbf{y}_{-i} ) \text{d} \boldsymbol{\theta} .
\end{nalign}
This integral has the form of equation~(\ref{eq:integ}) where the function $h$ is now the $i$'th likelihood term $p (y_i \mid \boldsymbol{\theta})$
and the probability distribution $p$ is
the corresponding $i$'th LOO posterior distribution $p (\boldsymbol{\theta} \mid \mathbf{y}_{-i} )$.
\citet{krueger2016probabilistic} prove that model assessment with the logarithmic score utility
is consistent when increasing the size of the posterior sample when using a Monte Carlo approximation to the posterior predictive distribution and a posterior sample generated using
a stationary and ergodic Markov chain.
They state that the theoretical conditions for the rate of convergence are difficult to verify.
Therefore, the Pareto diagnostics are important for monitoring the reliability of
model assessment.

Computing each of the $n$ integrals in equation~(\ref{eq:loopd})
using the simple Monte Carlo estimator is expensive because it
requires refitting the model $n$ times.
However,
if the observations are modeled as conditionally independent given the parameters $\boldsymbol{\theta}$ of the model, the likelihood factorizes as
\begin{nalign*}
	p(\mathbf{y} \mid \boldsymbol{\theta}) = \prod_{i = 1}^n p(y_i \mid \boldsymbol{\theta}) 
\end{nalign*}
and the LOO predictive density can be estimated with (self-normalized) importance sampling from the full data posterior~\citep{gelfand1992model}.
Here, we assume that only unnormalized posterior densities are available, and
present only the self-normalized importance sampling equations.
With draws $\{ \boldsymbol{\theta}^{(s)} \}_{s=1}^{S} $ from the full data posterior distribution $p (\boldsymbol{\theta} \mid \mathbf{y})$,
the unnormalized importance weights for the $i$'th LOO fold are defined as
\begin{nalign} \label{eq:rawisw}
w_{\text{loo},i}^{(s)} = \frac{1}{p (y_i \mid \boldsymbol{\theta}^{(s)})} \propto \frac{p (\boldsymbol{\theta}^{(s)} \mid \mathbf{y}_{-i} )}{p (\boldsymbol{\theta}^{(s)} \mid \mathbf{y} )} .
\end{nalign}
The self-normalized importance sampling estimator of equation~(\ref{eq:loopd}) is
\begin{nalign} \label{eq:isloosimp}
p (y_i \mid \mathbf{y}_{-i}) \approx  \frac{ \frac{1}{S} \sum_{s = 1}^{S} w_{\text{loo},i}^{(s)} \;  p (y_i \mid \boldsymbol{\theta}^{(s)})   }{ \frac{1}{S} \sum_{s = 1}^{S} w_{\text{loo},i}^{(s)}} =  \frac{1}{ \frac{1}{S} \sum_{s = 1}^{S} w_{\text{loo},i}^{(s)}}.
\end{nalign}

LOO-CV using the full data posterior as proposal distribution and
the log predictive density utility is a very special application
of self-normalized importance sampling for two reasons.
First, using the same proposal distribution for all LOO folds reduces the computational cost roughly by a factor equal to the number of observations
compared to directly sampling from each LOO posterior distribution. This is because
inference on the full data posterior and each LOO posterior is approximately equally expensive.
Second, the numerator of equation~(\ref{eq:isloosimp}) evaluates to one, which
indicates that the full data posterior is an optimal proposal distribution in terms of estimating the
numerator of the self-normalized importance sampling estimator.
Thus, only an adaptation targetting the denominator is required, whereas usually with
self-normalized importance sampling, two separate adaptations are required.
This is a good justification for using the full posterior as the proposal distribution
instead of a simpler parametric distribution.

\subsection{Implementing the Proposed Methods for Leave-One-Out Cross-Validation}

Here, we show the implementation of the importance weighted moment matching for leave-one-out cross-validation.
We focus on the case of self-normalized importance sampling with a sample from the full data posterior
distribution.
When sampling from the full data posterior $p (\boldsymbol{\theta}\mid \mathbf{y})$, the unnormalized common importance weights are given
by equation~(\ref{eq:rawisw}).
After an affine transformation, the importance weights are computed as
\begin{nalign} \label{eq:newIR}
\transf{w}_{\text{loo},i}^{(s)} = \frac{p (\transf{\boldsymbol{\theta}}^{(s)} \mid \mathbf{y})}{p(\boldsymbol{\theta}^{(s)}\mid \mathbf{y}) p(y_{i}\mid\transf{\boldsymbol{\theta}}^{(s)})} \propto \left ( \frac{p (\transf{\boldsymbol{\theta}}^{(s)} \mid \mathbf{y}_{-i})}{p(\boldsymbol{\theta}^{(s)}\mid \mathbf{y})} \right ).
\end{nalign}
While the denominator term $p (\boldsymbol{\theta}^{(s)}\mid \mathbf{y})$ is a constant for the $s$'th draw and equal for all LOO folds,
the additional cost compared to equation~(\ref{eq:rawisw}) is that for each transformed draw $\transf{\boldsymbol{\theta}}^{(s)}$,
both the full data posterior density $p (\transf{\boldsymbol{\theta}}^{(s)}\mid \mathbf{y})$ and
the likelihood term $p(y_{i}\mid\transf{\boldsymbol{\theta}}^{(s)})$
need to be evaluated, instead of just the likelihood.
However, even with multiple iterations, this cost is much
smaller than running a full inference on the LOO posterior.

After moment matching,
the transformations are combined as $T_w (\boldsymbol{\theta}) = T_{wm} ( ... T_{w2} ( T_{w1} (\boldsymbol{\theta})))$, and
only half of the
$S$ of the original draws $\{ \boldsymbol{\theta}^{(s)} \}_{s = 1}^S$ are transformed using $T_w (\boldsymbol{\theta}^{(s)})$:
\begin{nalign*}
1 \leq s \leq \frac{S}{2}: \quad  & \transf{\boldsymbol{\theta}}^{(s)}  =  T_w( \boldsymbol{\theta}^{(s)}) \\
\frac{S}{2} < s \leq S: \quad & \transf{\boldsymbol{\theta}}^{(s)}  =  \boldsymbol{\theta}^{(s)} .
\end{nalign*}
We construct analogically an inverse transformation $T_w^{-1} (\boldsymbol{\theta}) = T_{w1}^{-1} ( T_{w2}^{-1} ... ( T_{wm}^{-1} (\boldsymbol{\theta})))$ and a pseudo-set of draws as $\transf{\boldsymbol{\theta}}^{(s)}_{\text{inv}} = T_w^{-1} (\transf{\boldsymbol{\theta}}^{(s)})$, i.e.
\begin{nalign*}
1 \leq s \leq \frac{S}{2}: \quad  & \transf{\boldsymbol{\theta}}^{(s)}_{\text{inv}}  =   \boldsymbol{\theta}^{(s)} \\
\frac{S}{2} < s \leq S: \quad & \transf{\boldsymbol{\theta}}^{(s)}_{\text{inv}}  = T_w^{-1} ( \boldsymbol{\theta}^{(s)} ).
\end{nalign*}
Then, the importance weights are computed as
\begin{nalign*}
\transf{w}_{\text{loo, split},i}^{(s)} = & \frac{p(\transf{\boldsymbol{\theta}}^{(s)}\mid\mathbf{y}_{-i})} {     g_{\text{split,loo}} (\transf{\boldsymbol{\theta}}^{(s)})  }
 = \frac{p (\transf{\boldsymbol{\theta}}^{(s)} \mid \mathbf{y})}{g_{\text{split,loo}} (\transf{\boldsymbol{\theta}}^{(s)}) p(y_{i}\mid\transf{\boldsymbol{\theta}}^{(s)})} ,
\end{nalign*}
where $g_{\text{split,loo}} (\boldsymbol{\theta})$ is the split proposal distribution
\begin{nalign} \label{eq:splitprop}
g_{\text{split,loo}} (\boldsymbol{\theta}) \propto    p (\boldsymbol{\theta} \mid \mathbf{y})  +  p_{T_w} (\boldsymbol{\theta} \mid \mathbf{y}) \propto    p (\boldsymbol{\theta} \mid \mathbf{y})  +   |\mathbf{J}_{T_w}|^{-1}  p ({T_w}^{-1} (\boldsymbol{\theta}) \mid \mathbf{y}) .
\end{nalign}

In addition to the log likelihood values for each observation and
each posterior draw that are required by self-normalized importance sampling LOO-CV,
the user must now also provide functions for computing the log posterior density of the model
and the log likelihood based on parameter values in the unconstrained parameter space. The latter is required because moment matching in a constrained space via affine transformations might violate the constraints. Thus, the algorithm operates in the unconstrained space where each parameter can have any real value.
For example, model parameters that are constrained to be positive, can be unconstrained by a log-transformation.
The full method is presented in Algorithm~\ref{alg:mmloo}.

The moment matching method presented in this work is implemented in R~\citep{rlang}
so that users can easily compare the predictive performance
of models.
The complete code
is available on Github (\url{https://github.com/topipa/iter-mm-paper}).
The method is also implemented in the \texttt{loo} R package~\citep{loo}
for importance sampling LOO-CV.
We also provide convenience functions that implement the moment matching method 
for models fitted with probabilistic programming language 
Stan~\citep{carpenter2017stan}. In this case, it is enough that
the user supplies a Stan fit object, where the log likelihood
computation is included in the generated quantities block.
Internally, the method then uses the \texttt{loo} package for importance sampling,
and the given Stan fit object for computing the likelihoods
and posterior densities. Our code is specifically modularized to make it 
straightforward to implement the moment matching also 
for other fitted model objects.

\begin{algorithm*}[htb]
	\caption{\em adaptive moment matching for LOO-CV}\label{alg:mmloo}
	\begin{algorithmic}[1]
	    \STATE Define stopping threshold $k_{\text{threshold}}$ corresponding to Pareto $\hat{k}$
	    diagnostic value;
		\STATE Run inference to obtain a sample $\{ \boldsymbol{\theta}^{(s)} \}_{s = 1}^S$ from the full data posterior of the model $p (\boldsymbol{\theta} \mid \mathbf{y})$;
		\STATE For each draw $\boldsymbol{\theta}^{(s)}$, precompute the full data posterior density $Q_s = p (\boldsymbol{\theta}^{(s)}\mid\mathbf{y})$
		 
		\FOR{observation $i$ in $1:n$}
		\STATE Initialize draws for this LOO fold as $\{ \boldsymbol{\theta}_i^{(s)} \}_{s = 1}^S = \{ \boldsymbol{\theta}^{(s)} \}_{s = 1}^S$;
		\STATE Compute common importance weights $w_{\text{loo},i}^{(s)} = p (y_i \mid \boldsymbol{\theta}^{(s)})^{-1}$;
		\STATE Fit generalized Pareto distribution to the largest weights $w_{\text{loo},i}^{(s)}$ and report the shape parameter $\hat{k}_i$;
		\IF{   $\hat{k}_i < k_{\text{threshold}}$  }
		\STATE Compute the estimate $\widehat{\text{elpd}}_{\text{loo},i}$ using self-normalized importance sampling;
		\ELSE
		
		\STATE Run \textbf{Algorithm 2}: Moment matching for self-normalized importance sampling;

		\IF{   $\hat{k}_i < k_{\text{threshold}}$  }
		\STATE Compute the estimate $\widehat{\text{elpd}}_{\text{loo},i}$ using self-normalized importance sampling;
		\ELSE
		\STATE Run inference to obtain a sample $\{ \boldsymbol{\theta}_i^{(s)} \}_{s = 1}^S$ from the LOO posterior $p (\boldsymbol{\theta} \mid \mathbf{y}_{-i})$;
		\STATE Fit generalized Pareto distribution to the largest expectation-specific weights
		$\ome_{\text{loo},i}^{(s)} = p(y_i \mid \boldsymbol{\theta}^{(s)})$ and report the shape parameter $\hat{k}_i$;
		\IF{   $\hat{k}_i < k_{\text{threshold}}$  }
		\STATE Compute the estimate $\widehat{\text{elpd}}_{\text{loo},i}$ using simple Monte Carlo sampling;
		\ELSE

		\STATE Run \textbf{Algorithm 3}: Moment matching for simple Monte Carlo sampling;
		\IF{   $\hat{k}_i < k_{\text{threshold}}$  }
		\STATE Compute the estimate $\widehat{\text{elpd}}_{\text{loo},i}$ using simple Monte Carlo sampling;
        \ELSE
		\STATE Give a warning that estimating $\widehat{\text{elpd}}_{\text{loo},i}$ is difficult, and more Monte Carlo draws may help;
		\ENDIF
		
		\ENDIF
		\ENDIF
		\ENDIF
		\ENDFOR
		 
	\end{algorithmic}
\end{algorithm*}

\section{Additional Results}
\label{appendix:results}

\subsection{Normal Model: Optimality of the Split Proposal Distribution} \label{sec:optimality}

\begin{figure}[t]
\centering
\includegraphics[]{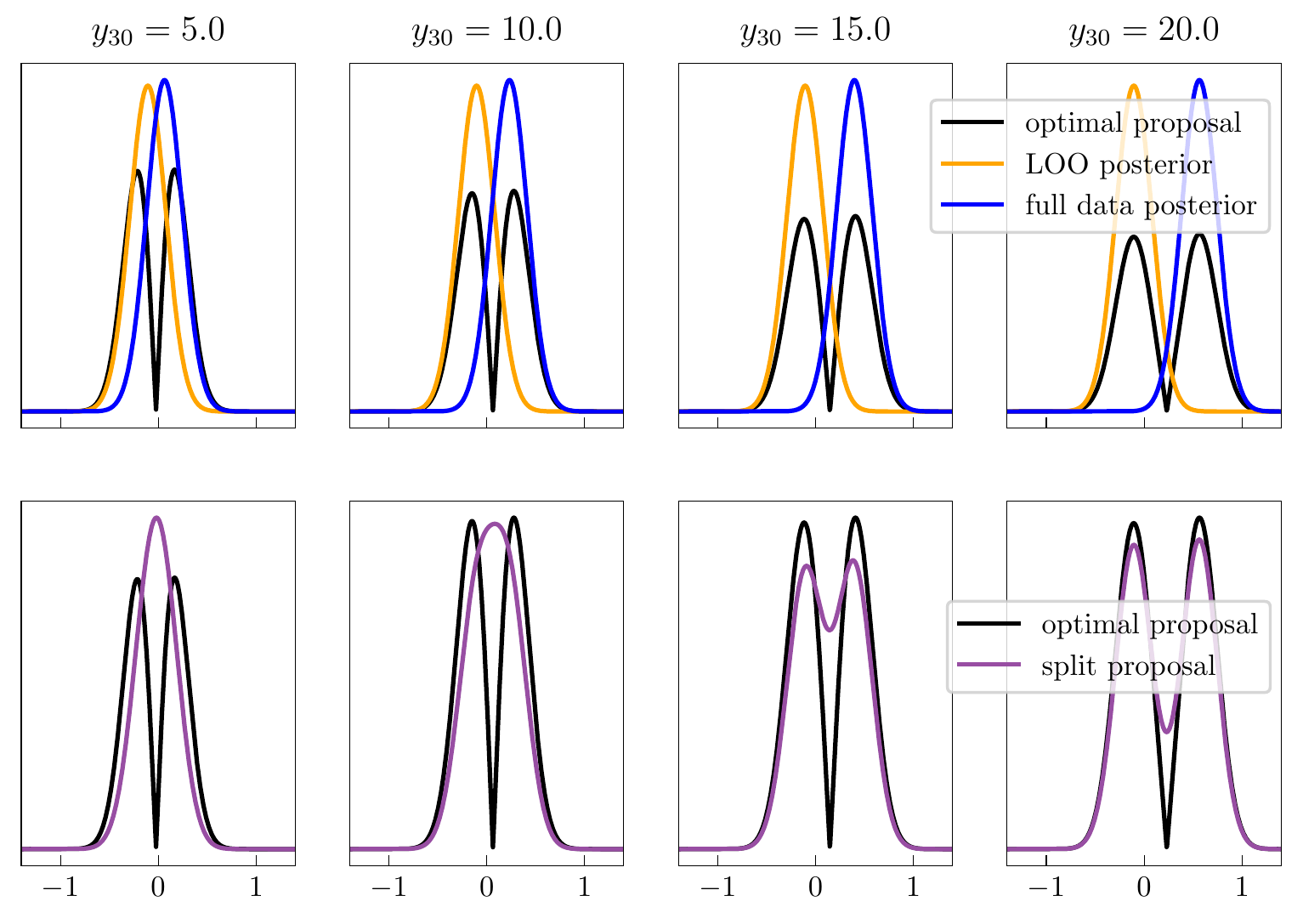}
\caption{For the normal model with known variance, the shape of
the optimal proposal distribution together with different proposal distributions
for different values of the outlier $y_{30}$.
Top row: LOO posterior and full data posterior.
Bottom row: Split proposal distribution from equation~(\ref{eq:5050prop}).} \label{fig:optimal_vs_smm_3}
\end{figure}

For illustrationary purposes, let us simplify the normal model from Section~\ref{sec:toyexperim} such that we assume the variance
of the normally distributed data is known. Then, the model has just one parameter, the mean
of the data, and the posterior distribution
of that parameter is Gaussian.
Using the one-dimensional posterior, we can efficiently visualize
why both the LOO posterior and the full data posterior can
be inadequate proposal distributions for self-normalized importance sampling LOO-CV.
In the top row of Figure~\ref{fig:optimal_vs_smm_3} we illustrate
the LOO posterior and the full data posterior of the model together with
the optimal proposal distribution for computing the self-normalized
importance sampling LOO-CV estimate when we move the outlier $y_{30}$ further. It is evident that
when the left-out observation is influential,
neither the LOO posterior nor the full data posterior can
provide enough draws from one of the tails to adequately estimate the LOO-CV integral.
In the bottom row of Figure~\ref{fig:optimal_vs_smm_3} we illustrate the split proposal distribution in equation~(\ref{eq:5050prop}), which
conversely becomes closer and closer to the optimal proposal distribution
when the left-out observation $y_{30}$ becomes more influential.

\subsection{Normal Model: Randomly Generated Data}

\begin{figure}[t]
\centering
\includegraphics[width=\textwidth]{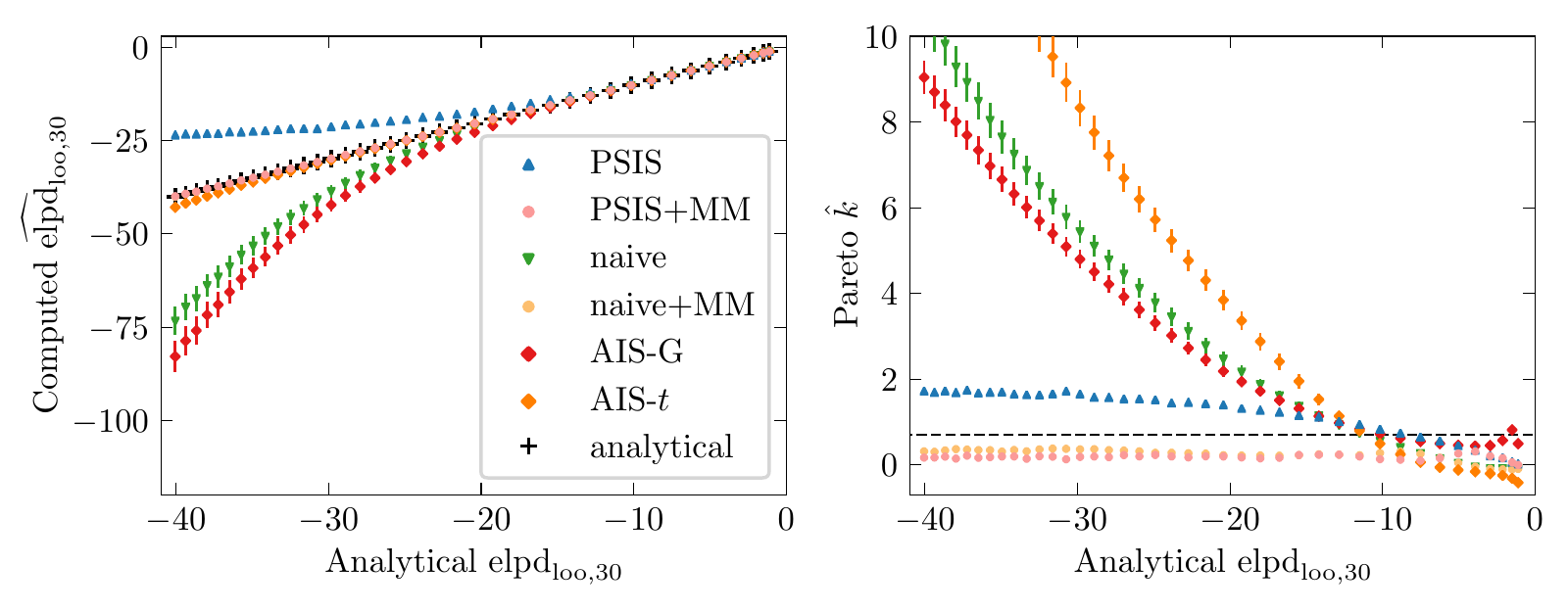}
\caption{Computed log predictive density estimates of the left out observation $y_{30}$
for different values between $y_{30} = 0$ and $y_{30} = 20$ in the Gaussian model of Section~\ref{sec:toyexperim}.
The black crosses depict the analytical LOO predictive density. The sampling results are averaged from 100 independent Stan runs, and the error bars represent $95 \%$
intervals of the mean across these runs. For every Stan run, the observations $y_{1}$ to $y_{29}$
are randomly re-generated.} \label{fig:toynormal_randomized}
\end{figure}

In Figure~\ref{fig:toynormal_randomized}, the results of Figure~\ref{fig:toynormal}
are replicated, but now the normally distributed observations $y_{1}$ to $y_{29}$
are different for each Stan run. The results are in principle similar to those discussed in
Section~\ref{sec:toyexperim}.

\subsection{Importance weighted moment matching without Pareto smoothing}

In Table~\ref{tab:results_raw}, we show
similar results as in Table~\ref{tab:results}, but
the importance weighted moment matching does not use Pareto smoothing to
smooth the importance weights during adaptation~\citep{vehtari2015pareto}.

\begin{table*}[tb]
\centering
\caption{Upper part: Numbers of LOO folds with Pareto $\hat{k}$ diagnostic above 0.7 when the models are fitted to the full data set (lower is better).
Lower part: Average run times in seconds for different algorithms.
Column PSIS corresponds to using the full data posterior directly
as the proposal distribution.
Columns PSIS+MM and IS+MM correspond to importance weighted
moment matching with an without Pareto smoothed importance weights, respectively.}
\label{tab:results_raw}
\begin{tabular}{ c }
\hspace{3cm} \textbf{Folds with} $\hat{k} > 0.7$ \hspace{3cm} \\
\end{tabular}
\begin{tabular}{ l r r r r }
\toprule
Data and model & Draws & PSIS & PSIS+MM & IS+MM \\
\midrule

Section~\ref{sec:poisson}:         & $2000$  & 15.2 & 0 & 0  \\
Roach data                         & $4000$  & 14.7 & 0 & 0 \\
Poisson regression model           & $8000$  & 14.2 & 0 & 0 \\

\hline                                    

Section~\ref{sec:linreg}:        & $2000$  & 14.0 & 0 & 0.5  \\
Correlated Predictor Variables                              & $4000$  & 13.8 & 0 & 0.3  \\
Linear regression model                                 & $8000$  & 13.4 & 0 & 0.2 \\

\hline                                    
Section~\ref{sec:ovarian}:        & $1000$  & 34.8 & 20.1 & 20.7  \\
Ovarian cancer data ($n < p$)               & $2000$  & 36.1 & 19.6 & 19.9  \\
Logistic regression model         & $4000$  & 34.8 & 16.2 & 17.1  \\
                                  & $8000$  & 34.0 & 11.4 & 13.5 \\

\bottomrule

\end{tabular}

\begin{tabular}{ c }
\, \\
\textbf{Computation times (seconds)}  \\
\end{tabular}

\begin{tabular}{ l r r r r }
\toprule
Data and model & Draws & PSIS & PSIS+MM & IS+MM \\
\midrule

Section~\ref{sec:poisson}:         & $2000$  & 0 & 39 & 39  \\
Roach data                         & $4000$  & 0 & 70 & 70 \\
Poisson regression model           & $8000$  & 0 & 140 & 140 \\
\hline     

Section~\ref{sec:linreg}        & $2000$  & 0 & 52 & 51    \\
Correlated Predictor Variables                              & $4000$  &  0 & 114 & 114   \\
Linear regression model   & $8000$  & 0  & 346 & 340  \\
   
   \hline
   
Section~\ref{sec:ovarian} & $1000$  & 0 & 199 & 181  \\
Ovarian cancer data ($n < p$)             & $2000$  & 0 & 372 & 344  \\
Logistic regression model         & $4000$  & 0 & 1558 & 1566   \\
      & $8000$  & 0 & 3733 & 3816   \\
   
\bottomrule

\end{tabular}
\end{table*}

\end{appendices}

\end{document}